\documentclass[sigconf, nonacm, pdfa]{acmart}
 
\newcommand\vldbdoi{10.14778/3750601.3750635}
\newcommand\vldbpages{5166 - 5183}

\newcommand\vldbvolume{18}
\newcommand\vldbissue{12}
\newcommand\vldbyear{2025}

\newcommand\vldbauthors{\authors}
\newcommand\vldbtitle{\shorttitle} 

\newcommand\vldbavailabilityurl{URL_TO_YOUR_ARTIFACTS}

\usepackage[a-2b]{pdfx}
\usepackage{todonotes}
\usepackage{enumitem}
\usepackage{algpseudocode}
\usepackage{pseudocode}
\usepackage[ruled,vlined]{algorithm2e}
\usepackage{pifont}
\usepackage{listings-rust}
\usepackage{xcolor}
\usepackage{subcaption, graphicx}
\usepackage{url}
\usepackage{balance}
\usepackage{multirow}

\definecolor{codegreen}{rgb}{0,0.6,0}
\definecolor{codegray}{rgb}{0.5,0.5,0.5}
\definecolor{codepurple}{rgb}{0.58,0,0.82}
\definecolor{backcolour}{rgb}{0.97,0.97,0.97}
\lstdefinestyle{mystyle}{
    backgroundcolor=\color{backcolour},   
    commentstyle=\color{codegreen},
    keywordstyle=\color{magenta},
    numberstyle=\tiny\color{codegray},
    stringstyle=\color{codepurple},
    basicstyle=\ttfamily\footnotesize,
    breakatwhitespace=false,         
    breaklines=true,                 
    captionpos=b,                    
    keepspaces=true,                 
    numbersep=5pt,                  
    showspaces=false,                
    showstringspaces=false,
    showtabs=false,                  
    tabsize=2
}

\newcommand{\dbpoint}[1]{\ensuremath{{\sf x}_{#1}}}
\newcommand{\nout}[1]{N_{\rm out}({#1})}

\newcommand{\greedysearch}[4]{\ensuremath{{\rm GreedySearch}({#1},{#2},{#3},{#4})}}
\newcommand{\betasearch}[6]{\ensuremath{{\rm BetaSearch}({#1},{#2},{#3},{#4},{#5},{#6})}}

\newcommand{\prune}[4]{\ensuremath{{\rm RobustPrune}({#1},{#2},{#3},{#4})}}

\newcommand{\insertpoint}[5]{\ensuremath{{\rm Insert}({#1},{#2},{#3},{#4},{#5})}}
\newcommand{\batchinsertpoint}[5]{\ensuremath{{\rm MiniBatchInsert}({#1},{#2},{#3},{#4},{#5})}}

\newcommand{\cL}{\mathcal{L}\xspace}
\newcommand{\cV}{\mathcal{V}\xspace}
\newcommand{\cE}{\mathcal{E}\xspace}

\newcommand{\hops}{\ensuremath{\mathsf{hops}}\xspace}
\newcommand{\cmps}{\ensuremath{\mathsf{cmps}}\xspace}

\newcommand{\diskann}{{\ensuremath{\rm DiskANN}}\xspace}

\newcommand{\cosmosdb}{Cosmos DB\xspace}
\newcommand{\fflat}{Flat\xspace}
\newcommand{\qflat}{Quantized Flat\xspace}
\newcommand{\bwtree}{Bw-Tree\xspace}

\newcommand{\magdalen}[1]{}
\newcommand{\harsha}[1]{}

\begin{document}
\title{Cost-Effective, Low Latency Vector Search with Azure \cosmosdb}

\author{Nitish Upreti,
Harsha Vardhan Simhadri,
Hari Sudan Sundar,
Krishnan Sundaram,
Samer Boshra,
Balachandar Perumalswamy,
Shivam Atri,
Martin Chisholm,
Revti Raman Singh,
Greg Yang,
Tamara Hass,
Nitesh Dudhey,
Subramanyam Pattipaka,
Mark Hildebrand,
Magdalen Manohar,
Jack Moffitt,
Haiyang Xu,
Naren Datha,
Suryansh Gupta,
Ravishankar Krishnaswamy,
Prashant Gupta,
Abhishek Sahu, 
Hemeswari Varada,
Sudhanshu Barthwal,
Ritika Mor,
James Codella,
Shaun Cooper,
Kevin Pilch,
Simon Moreno,
Aayush Kataria,
Santosh Kulkarni,
Neil Deshpande,
Amar Sagare,
Dinesh Billa,
Zishan Fu,
Vipul Vishal 
}
\affiliation{\institution{Microsoft}}
\email{{niupre, harshasi, harsudan}@microsoft.com}

\begin{abstract}

Vector indexing enables semantic search over diverse corpora and 
has become an important interface to databases for both users and AI agents. 
Efficient vector search requires deep optimizations in database systems.
This has motivated a new class of specialized vector databases that optimize 
for vector search quality and cost. Instead, we argue that a scalable,
high-performance, and cost-efficient vector search system can  be built 
inside a cloud-native operational database like Azure \cosmosdb{} while leveraging the
benefits of a distributed database such as high availability, durability, and scale.
We do this by deeply integrating DiskANN, a state-of-the-art vector indexing library,
inside  Azure \cosmosdb NoSQL.
This system uses a single vector index per partition stored in existing index trees,
and  kept in sync with underlying data. 
It supports < 20ms query latency over an index spanning 10 million vectors,
has stable recall over updates, and offers approximately $43\times$ and $12\times$ lower query cost compared to Pinecone and Zilliz serverless enterprise products.
It also scales out to billions of vectors via automatic partitioning.
This convergent design presents a point in favor
of integrating vector indices into operational databases in the context
of recent debates on specialized vector databases, and offers a
template for vector indexing in other databases.
\end{abstract}

\maketitle

\pagestyle{empty}
\begingroup\small\noindent\raggedright\textbf{PVLDB Reference Format:}\\
\vldbauthors. \vldbtitle. PVLDB, \vldbvolume(\vldbissue): \vldbpages, \vldbyear.\\
\href{https://doi.org/\vldbdoi}{doi:\vldbdoi}
\endgroup
\begingroup
\renewcommand\thefootnote{}\footnote{\noindent
This work is licensed under the Creative Commons BY-NC-ND 4.0 International License. Visit \url{https://creativecommons.org/licenses/by-nc-nd/4.0/} to view a copy of this license.
For any use beyond those covered by this license, obtain permission by emailing \href{mailto:info@vldb.org}{info@vldb.org}. Copyright is held by the owner/author(s). Publication rights licensed to the VLDB Endowment. \\
\raggedright Proceedings of the VLDB Endowment, Vol. \vldbvolume, No. \vldbissue\ %
ISSN 2150-8097. \\
\href{https://doi.org/\vldbdoi}{doi:\vldbdoi} \\
}\addtocounter{footnote}{-1}\endgroup


\section{Introduction}
Advances in deep learning have made it possible to {\it embed}
data as vectors in high-dimensional vector spaces so that 
the distance between vectors captures various notions of
semantic similarity. 
This opens up a new interface for users as well as AI models and agents
to interact with large information stores
based on semantic and contextual relevance,
rather than literal matches or structured queries.
%
%
Therefore, efficient search in vector spaces has become a critical 
requirement for information retrieval systems.
Already, vector search is a central component in industrial scale
retrieval (web and document search) and recommendation systems.
In databases, especially document databases, augmenting existing workloads
with vector representations is becoming commonplace.

A new class of scenarios requiring vector search on operational data 
modeling e-commerce, document retrieval, conversational histories,
and AI agent interaction patterns, are rapidly growing.
This has motivated a new class of specialized vector databases
that optimize primarily for vector search performance.
However, this pattern forces the replication of data between 
a primary operational database and a secondary vector database
which can cause data divergence, and increased cost and complexity
for the user.
This also may not provide the operational resilience that developers expect.

An ideal solution to these workloads would be a highly available and scalable 
operational database that allows flexible data models and indexing over vector representations.
It would further offer:
\begin{itemize}[topsep=0pt,noitemsep, leftmargin=*]
\item A vector index in sync with underlying data without replication to external systems.
\item Elastic scaling to billions of vectors with thousands of dimensions. 
\item Cost-effective and accurate search at any scale and QPS.
\item Robustness to incremental changes -- ensures data integrity and consistency, 
    and high search accuracy across updates. 
\item Low-latency transactions for data updates and retrieval. 
\item Built-in multi-tenancy to allow multiple users or groups to securely and cost-effectively share the same database instance. 
\end{itemize}


We achieve all these properties by {\bf integrating a state-of-the-art vector
indexing library, DiskANN, with Azure \cosmosdb for NoSQL},
an operational database for Tier-0 enterprise workloads.
\cosmosdb already stores vast quantities of data such as conversational history,
documents and e-commerce data where semantic retrieval is important.
The database engine underlying \cosmosdb NoSQL~\cite{AzureDocDB}
already offers many features to help realize these properties
including multi-tenant support, automatic indexing with flexible schema,
 scale-out and high-availability architecture, multi-region support,
as well as flexible cost structures such as dynamic auto-scale and pay-per-use serverless models.
We take advantage of these properties by adapting the DiskANN
library within the contours of the \cosmosdb architecture.


Each collection in \cosmosdb maps to multiple physical partitions (based on hashed key ranges), each of which is made highly available with a replica-set.
Each physical machine in a \cosmosdb cluster hosts partitions corresponding to many
collections to maximize fleet efficiency. Therefore,
only a portion of available memory in these machines is available
as a cache for the indices in these replicas. 
The available cache might be $10-50\times$
lower than the size of the document data and indices over them.
For a vector index to be effective and cost-efficient in such a constrained setting,
it must be able to process incremental updates and queries with limited memory.
Moreover, the index must be truly incremental and avoid needing to be rebuilt or merged
to maintain search quality over a long period of time or a large number of operations.

DiskANN is a suite of vector indexing algorithms~\cite{tode25, diskann19, freshdiskann}
designed for such  constraints. It derives \emph{quantized} representations
of vectors which can be much smaller than
the original vectors,
and supports updates and queries to the index mostly via quantized vectors.
To perform a query, full-precision vectors stored in
the index are accessed $50\times$ less frequently compared to quantized vectors. 
This allows the system to provide high performance even when most of the index
is stored on SSDs. DiskANN has been widely deployed
in several extreme scale: Microsoft semantic indices used for web search,
advertisements, Microsoft 365 search, Co-pilots as well as on edge devices~\cite{windowscopilot}.

The DiskANN library~\cite{tode25} has previously been designed
to control the layout of index terms either in memory or in SSD,
akin to other monolithic systems ~\cite{Faiss17, HNSW-github}.
While databases such as SingleStore~\cite{singlestore-v}
and Elastic~\cite{elasticsearch-vector, elasticsearch-vector2}
use vector indexing libraries in a loosely coupled way 
to produce a separate index for each immutable data segment, 
we do not use such a design due to several drawbacks:
(a) each query has to fan out to
numerous segments which limits query efficiency, 
(b) regular rebuilds of vector indices are needed as segments are merged or consolidated,
which consumes significant compute, memory
and causes serious latency spikes for queries~\cite{lucene-hnsw-cpu, lucene-hnsw-memory},
(c) either large cold start latencies occur while loading large vector indices into memory,
or high expense is incurred from having to hold the index in memory.

Instead, for simplicity and robustness of the system, we store the
terms representing the vector index on the \bwtree index in \cosmosdb.
\bwtree supports  concurrency through latch free algorithms,
provides fast random reads and writes in a tiered memory/SSD setting. 
To enable this, we rewrote the DiskANN library to decouple
the algorithmic logic from physical index layout. The new library
provides supports updates and queries by manipulating or reading index terms
(quantized vector and graph adjacency lists)
stored external to the library.

This leads to several structural advantages:
\begin{itemize}[leftmargin=*, noitemsep]
\item We can maintain and update just one vector index per replica, which can be as large as 50GB, enabling higher query efficiency via reduced fan-out. In fact, DiskANN's query complexity scales logarithmically (empirically) in the size of the index.
\item   A vector insert results in immediate and durable changes to the index 
terms to  Bw-Tree, and does not require further indexing or merging down the line.
\item The \bwtree caches index terms for hot partitions on demand, while cold collections
do not consume memory.
\item A long tail of indices can be stored in a machine 
without paying a minimum floor cost, especially in multi-tenant collections.
\end{itemize}

This deep integration eliminates the need for a vector index
outside an operational database, and instead composes proven features from \cosmosdb and DiskANN.
From \cosmosdb, we inherit flexible cost structures, resource governance, elasticity, and resiliency. 
By storing its indexing terms in \bwtree, which has a long history of support in \cosmosdb, 
DiskANN  benefits from its established stability and ease of operation.
From the new rewrite of DiskANN we inherit existing features -- querying with limited memory
-- as well as new features developed for this integration such as index updates
with limited memory, and filter-aware and paginated search for efficiently processing
hybrid vector queries with predicates.

A few highlights of the integrated system include:
\begin{itemize}[noitemsep, leftmargin=*]
\item Query cost that is nearly $12\times$ and $43\times$ lower than Zilliz and Pinecone enterprise-tier
serverless vector databases respectively (for 10 million 768D vectors), while offering higher availability. 
\item Query latency of about 20 milliseconds, including the time to fetch underlying docs,
even at 10 million index scale. 
\item Query cost increases less than $2\times$ despite a $100\times$ increase in index size in one partition.
Query cost does not change much  as the dimensionality of the vector increases. 
\item Ingest offers stable recall over long update sequences, with cost and performance comparable to other vector databases. 
\item Collections can scale out to a billion vectors.
\item Multi-tenant design where the number of partition keys or the
vectors per partition can grow independently to large numbers.
\item Pay-per-use or auto-scale cost structure.
\item Optimized hybrid queries for better latency and cost than paginated search
with post-filtering.
\end{itemize}


\section{background}\label{sec:background}
We now review necessary background on the DiskANN library and the \cosmosdb system
to motivate the new design  in  Section~\ref{sec:design}. 

\subsection{The DiskANN vector indexing library}\label{ref:diskannlibrary}
DiskANN is a graph-structured index for vector search that can
efficiently index and update large sets of vector data,
while supporting accurate and fast vector search queries. 
It is widely used at scale in Microsoft for semantic indices including those in
web search, enterprise document search, computational advertisement and Windows Co-pilot runtime.
The overall ideas are described in ~\cite{tode25} and an open-source implementation is available~\cite{diskann-github}.

The index consists of a graph over the vectors in the database,
with one vertex representing each vector and directed edges connecting vertices.
The search for a query $q$ uses a ``greedy'' approach, starting at a designated start point $s$, computing the distances from $q$ to each point in the out-edges $\nout{s}$, and moving on to the nearest neighbor of $q$ among $\nout{s}$. It continues this process of greedily visiting the closest neighbor to $q$ until it can no longer improve on the closest neighbor to $q$, at which point the search terminates. This algorithm is formally described in Algorithm~\ref{alg:greedysearch}, and can be naturally extended to return the top-$k$ neighbors by keeping a priority queue of the $k$ closest neighbors instead of hopping to the single closest neighbor each time. The accuracy, or \textit{Recall k@k}, of a search, is defined as how many of the $k$ results returned by a search are the true top-$k$ nearest neighbors. When clear from context, it is also used to refer to the average recall over a batch of searches.

DiskANN’s query complexity grows logarithmically with the size
of the index (empirically observed~\cite{diskann19, ParlayANN24}).
It can work effectively with an almost entirely SSD-based index
and limited memory,
while providing performance parity with in-memory indices
like ScaNN that consume an order of magnitude more memory.
DiskANN is IO efficient – it can achieve 90\% recall@10
on a billion-size SIFT dataset with just 50 random 4 KB reads to SSD.
The same DiskANN index can also be loaded entirely into DRAM for scenarios 
requiring extreme throughput, where it outperforms graph-based methods 
such as HNSW~\cite{hnswlib} and partition-based methods such as IVF and ScaNN~\cite{diskannwhitepaper}. 

\begin{algorithm}[t]
	\DontPrintSemicolon \small \KwData{ Graph $G$ with start node
          $s$, query \dbpoint{q}, result size $k$, search list size $L
          \geq k$}
	
	\KwResult{$\cL$ contains $k$-approx NNs, and set of visited nodes $\cV$}
	
	\tt
	\Begin{
		initialize sets $\cL\gets \{s\}$, $\cE\gets\emptyset$, and $\cV\gets\emptyset$\;
        \tcp{\small{$\cL$ is the list of best $L$ nodes, $\cE$ is the set 
        of nodes which have already been expanded from the list, 
        $\cV$ is the set of all visited nodes, i.e., inserted into the list}}
        initialize $\hops\gets 0$ and $\cmps \gets 0$\;
		\While{$\cL \setminus \cE \neq \emptyset$}{
			let $p* \gets \arg \min_{p \in \cL\setminus \cE} || \dbpoint{p} - \dbpoint{q}||$\;
			update $\cL \gets \cL \cup (\nout{p^*} \setminus \cV)$ and $\cE \gets \cE \cup \{p^*\}$\;
			\If{$|\cL| > L$} {
				update $\cL$ to retain closest $L$ points to \dbpoint{q}\;
			}
                update $\cV \gets \cV \cup \nout{p^*}$
		}
		return $[$closest $k$ points from $\cV$; $\cV$$]$\;
	}
	\caption{\greedysearch{s}{\dbpoint{q}}{k}{L}}
	\label{alg:greedysearch}
\end{algorithm}

\begin{algorithm}[t]
  \DontPrintSemicolon
  \small
  \KwData{Graph $G(P,E)$ with start node $s$, new vector $\dbpoint{p}$, parameter $\alpha > 1$, out degree bound $R$, list size $L$}
  \KwResult{Graph $G'(P',E')$ where $P' =  P \cup \{p\}$}
  \tt
  \Begin{
      initialize expanded nodes $\cE \gets \emptyset$\;
      initialize candidate list $\mathcal{L} \gets \emptyset$\;
      $[\mathcal{L},\cE] \gets \greedysearch{s}{p}{1}{L}$        \;
      set $\nout{p} \gets \prune{p}{\cE}{\alpha}{R}$  \;
      \ForEach{$j \in \nout{p}$}{
        \eIf{$|\nout{j} \cup \{p\}| > R$}
            {
              set $\nout{j} \gets \prune{j}{\nout{j} \cup \{p\}}{\alpha}{R}$\;
            }{
              update $\nout{j} \gets \nout{j} \cup \{p\}$\;
            }
      }
    }
    \caption{\insertpoint{\dbpoint{p}}{s}{L}{\alpha}{R}}
    \label{alg:insert}
\end{algorithm}

\begin{algorithm}[t]
	\DontPrintSemicolon \small \KwData{ Graph $G$, point $p \in
		P$, candidate set $\cE$, distance threshold $\alpha\geq 1$,
		degree bound $R$ } \KwResult{$G$ is modified by setting at
		most $R$ new out-neighbors for $p$}
	
	\tt
	\Begin{
		$\cE \gets (\cE \cup \nout{p}) \setminus \{p\}$\;
        $\cE \gets $ sort $\cE$ by distance from $p$ \;
		$\mathcal{N} \gets \emptyset$\;
        \For{$\dbpoint{q}, dist\_qp \in \cE$}{
            $add \gets \text{True}$ \;
            \For{$(\dbpoint{r}, dist\_rp \in \mathcal{N}$)}{
                \If{$\alpha*dist\_rp < ||\dbpoint{q}-\dbpoint{r}||$}{
                    $add \gets \text{False}$ \;
                    break \;
                }
            }
            \If{add}{
                $\mathcal{N} \gets \mathcal{N} \cup \{(\dbpoint{q}, dist\_qp) \}$ \;
            }
            
            \If{|$\mathcal{N}| = R$} {
				break\;
			}
        }

        $\nout{p} \gets \{q \text{ for } (\dbpoint{q}, dist\_qp) \in \mathcal{N}\}$ \;

	}
	\caption{\prune{p}{\cE}{\alpha}{R}}
	\label{alg:robustprune}
\end{algorithm}

\paragraph{Inserts and Replaces} The DiskANN graph is built via repeated calls to the insertion algorithm, which is formally described in Algorithm~\ref{alg:insert}. At a high level, the insert procedure generates candidates for insertion using a call to Algorithm~\ref{alg:greedysearch}, prunes the candidates down to respect the degree bound $R$, and then adds edges pointing to the newly inserted node to make it reachable. One of the 
main innovations behind DiskANN's performance is the RobustPrune routine, shown in Algorithm~\ref{alg:robustprune}, which is used to prune a vertex's out-neighbors down to the degree bound $R$. At a high level, it removes an edge $(u,v)$ when $v$ is likely to be reachable via one of $u$'s other neighbors. Furthermore, it utilizes a scaling parameter $\alpha$ to prune more or less aggressively; in practice, the ability to scale $\alpha$ to prune less aggressively is consequential for performance~\cite{diskann19,freshdiskann}. In some cases, it may also be necessary to \textit{replace} the vector corresponding
to a document identifier. We handle this by overwriting the original vector and invoking Algorithm~\ref{alg:insert} to re-insert the new vector. Any pre-existing edges pointing to the replaced point are cleaned up lazily via later calls to pruning.

\paragraph{Mini-batch updates} 
Using multiple threads to insert vectors in a DiskANN graph will result in 
race conditions between updates to the adjacency lists in the graph which could be handled via fine-grained locking~\cite{diskann-github}.
In some cases, the underlying data structure used to store the graph, may not tolerate parallel updates  / parallel updates with potential duplicate values for a key (graph vertex in this case).  The latter scenario also includes CosmosDB Bw-Tree with stricter contracts around no duplicate insert patches for key and delete patches for a non-existing key.
In order to benefit from parallelism while adhering to the above strict contracts, we utilize so-called mini-batch updates, where the edge insertions corresponding to a small batch of nodes are computed in parallel, then applied to the graph in a single update.
They are formally described in Algorithm~\ref{alg:minibatchinsert} in the Appendix, and follow a similar routine to the batch build in ParlayANN~\cite{ParlayANN24}. Here, we use a smaller maximum batch size (about 100) to support batch updates.

\paragraph{In-place Deletion} When a document is deleted,
the corresponding vector and quantized vector are immediately removed from the index.
Diskann removes the graph index terms to reflect the deletion
using Algorithm~\ref{alg:delete} in the Appendix, which is an adaptation of~\cite{xu2025inplace}.
At a high level, the algorithm replaces the critical connections to the deleted point, ensuring stability of the index quality.
A lightweight background process continuously removes any remaining edges pointing to the deleted point. Experiments show~\cite{xu2025inplace} that the combination of in-place deletion and lightweight background consolidation is effective at keeping recall stable over long cycles of insertion and deletion.

\paragraph{Compressing Vectors via Quantization}
An additional algorithmic building block is quantization to  compress the vectors.
This allows them to be stored more compactly in expensive storage tiers (e.g., main memory),
transmitted more efficiently across memory bus, and distance comparisons
to be computed with fewer CPU cycles with little loss in accuracy.
One popular method is scalar quantization which maps each coordinate of the
embedding to a smaller representation. For example, 32-bit floating point representations
are easily rounded to the nearest 16-bit floating point with little loss of precision. Rounding to
the nearest 1- or 2-bit representations is more lossy at a coordinate
level, but could still preserve enough information overall for the query to navigate the index.

Product quantization (PQ)~\cite{jegou2010product}  maps
collections of coordinates to a few bytes by clustering data and mapping
the data to the identity of the coordinate center. For
many datasets, product quantization achieves better compression than scalar quantization
normalized for errors introduced in distance calculations. For example, PQ can compress OpenAI’s
text-3-large embeddings (12KB) by 96x while retaining enough information to navigate the index. 
Although PQ was formulated for preserving euclidean distances between vectors,
in practice it also preserves inner product distances reasonably well.

\subsection{The \cosmosdb system}
 
Azure \cosmosdb is Microsoft’s globally distributed, elastic,
cloud-native database service for managing JSON documents at Internet scale. 
It is the primary database service in Microsoft cloud with 10s of millions of database partitions,
100+ PBs of data under management and 20M+ vCores.
Prior work presents a detailed description of the overall system~\cite{AzureDocDB}.
Here we briefly present ideas necessary for the design of the integrated vector index.

\paragraph{Schema Agnostic Indexing}
\cosmosdb uses the simplicity of JSON and its lack of a schema specification.
No assumptions are made about the documents stored and they can vary in schema.
\cosmosdb operates directly at the level of JSON grammar, 
blurring the boundary between the structure and instance values of documents.
This, in turn, allows the database to be "schema-free",
enabling it to automatically index documents without requiring schema or secondary indices.
For additional control, a custom indexing policy\cite{cosmosdb-indexingpolicy} can also be used to index specific properties using the 'path' notation, that allows precise navigation to specific substructures in a JSON document. For example, the property path '/employee/name' represents the 'name' node in the 'employee' object.
\cosmosdb, in addition to automatic indexing supported for the JSON type system, 
also supports specialized indexes such as spatial indexing, and more recently,
indexing for vector and full-text search. 
For faster query and to avoid storage bloat associated with JSON text,
\cosmosdb employs a custom binary encoding~\cite{cosmosdbdevblogbinaryencoding} for JSON data. 


\paragraph{Logical Partitioning and Elasticity}
Clients define a logical partition key on a Collection 
(up to three levels of hierarchy are allowed in partition keys~\cite{cosmosdb-elasticity-hierarchical-pk}).
A Collection can thus span multiple physical partitions, with data hashed
horizontally across partitions based on the logical partition key value.
As clients adjust throughput and/or storage needs, the compute
and storage required for collections are scaled out or scale back through partition splits and merges.

\begin{figure}
    \centering
        \includegraphics[width=\linewidth]{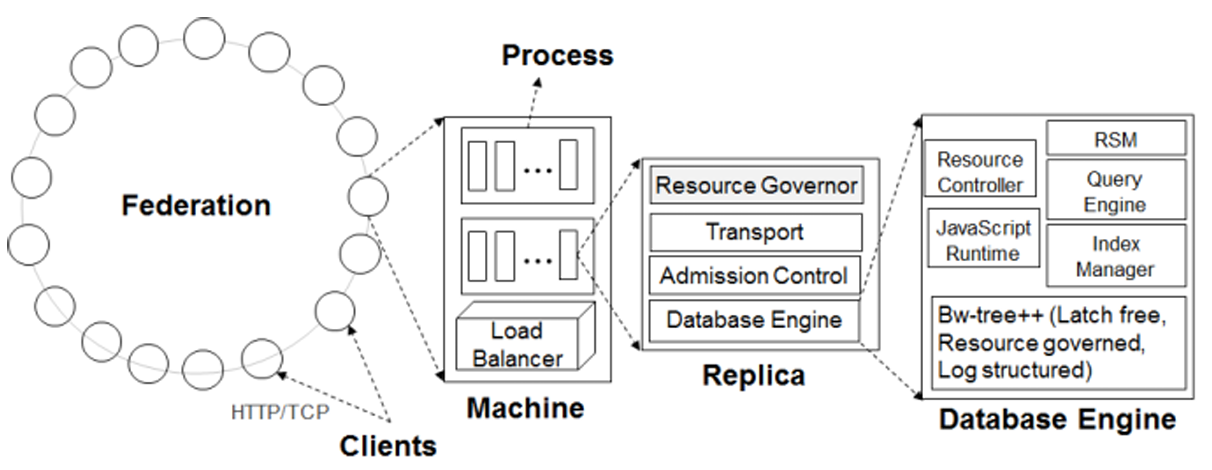} 
        \vspace{-10pt}
    \caption{Azure CosmosDB architecture diagram from~\cite{AzureDocDB}.}
    \Description[Azure CosmosDB architecture diagram]{Azure CosmosDB architecture diagram}
    \label{fig:cosmosdb-architecture}
\end{figure}

\paragraph{System Topology}
\cosmosdb service is deployed worldwide on clusters of machines each with dedicated local SSDs.
The unit of deployment, called a federation (Figure \ref{fig:cosmosdb-architecture}),
is an overlay network of machines, which can span one or more clusters. 
Each physical machine hosts replicas corresponding to various partitions for scaled out collections.
Replicas corresponding to a single partition are placed and load balanced across machines spanning different fault domains, upgrade domains and availability zones in the federation. Each replica hosts an instance of \cosmosdb database engine,
which manages the JSON documents as well as the associated indices. 


\paragraph{Resource Governance}
\cosmosdb 
provides performance isolation
between Replicas through the Resource Governance (RG) component.
Azure \cosmosdb normalizes the cost of all database operations using
Request Units (or RUs, for short) and measures cost based on throughput
(RUs per second, RU/s). Request unit is a performance currency
abstracting the system resources such as CPU, IOPS, and memory that are
required to perform the database operations.
The RG component guarantees the provisioned RUs
per partition for user requests while at the same time rate limiting requests
when usage exceeds provisioned throughput. This helps provide an isolated
provisioned throughput experience, while achieving high utilization per node and lower cost.


A Cosmos DB Replica's database engine hosts:

\begin{itemize}[noitemsep, leftmargin=*]
\item \textit{Document Store} - \cosmosdb's transactional database engine that serves as the main store for documents.

\item \textit{Inverted and Forward Index} - \cosmosdb incorporates Bw-Tree both
as an Inverted and Forward Index for its indexing needs. 
The Bw-Tree is a latch-free index that is designed for fast writes, 
thanks to its support for blind incremental updates and underlying log structured storage for persistence. The design effectively batches multiple incremental updates into a single flush on to disk. As a result, Bw-Tree does not update in-place,
helping it reduce write amplification to SSD based storage. \cite{AzureDocDB}

\end{itemize}

\vspace{-7pt}
\section{System Design}
\label{sec:design}

To create a vector index over a collection,
users turn on the capability at a collection level and 
declare a JSON path as the target for vector indexing.
In addition, users specify the dimension and the distance function
to be used for the embeddings in this path, as well as the vector indexing policy.
Any document ingested with a valid vector will be part of the vector index
(we support one vector per path).
The JSON document along with the vector is stored in the primary Document Store and
other paths meant for non-vector indexing are indexed into the \bwtree as described in~\cite{AzureDocDB}.

The simplest (brute force) way to compute the nearest neighbors
of a query is to scan all JSON documents in the collection in the query runtime,
and compute the distance to each vector in the designated vector path.
This is useful for small collections, say with less than 1000 documents but is not scalable otherwise.

An improvement would be to map all vectors in the collection
to a contiguous range in the \bwtree by prefixing the vectors
with the collection id and the name of the vector index path (we call this the \fflat index).
This can be done easily in the user request path as part of 
the "Document Analysis" during ingestion on the primary replica,
and then replicated to all secondaries.
A range scan in the \bwtree for the appropriate prefix 
would need many fewer random accesses for smaller vectors.
This is still not a great option since vectors tend to be large and storing them twice limits scale up. 

A second improvement  would be to compress the vectors
via quantization (say using PQ) and store them in a contiguous range in the \bwtree.
This significantly reduces the number of nodes to scan in the \bwtree,
by up to $96\times$ for OpenAI Ada v3 embeddings, for example.
To answer a request for the top-$k$ nearest neighbors to query $q$,
we would first find, say, $5k$ entries to the query in the quantized space.
We could then look up the full precision vectors corresponding to each of the
$5k$ candidates from Document Store, and compute the full fidelity distance to the query
to identify the top-$k$ candidates. We refer to this as the Q-Flat (\qflat) index.
With an appropriate multiplier over how many extra candidates are retrieved in 
quantized space, this method can yield very high recall.
For moderate sized collections, or for small tenants in a multi-tenant collection,
this can be efficient. For instance, in a collection of $5,000$ vectors quantized to $96$ bytes,
we need to touch about $60$ 8KB-sized \bwtree logical leaf nodes to answer the query
and use $<1ms$ of CPU time to  compute distances.

\begin{figure}
    \centering
    \includegraphics[width=0.48\linewidth]{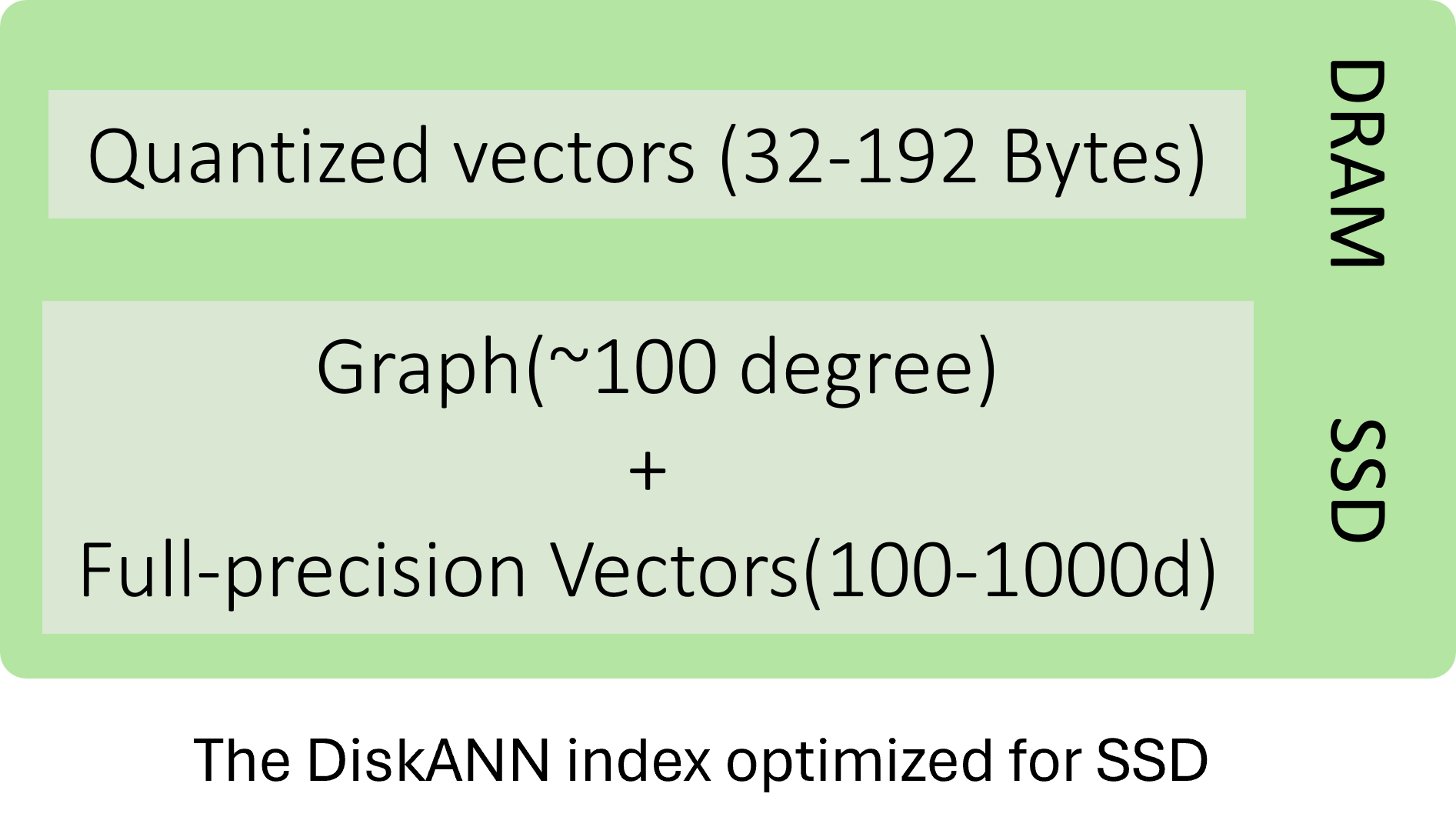}
~
    \includegraphics[width=0.51\linewidth]{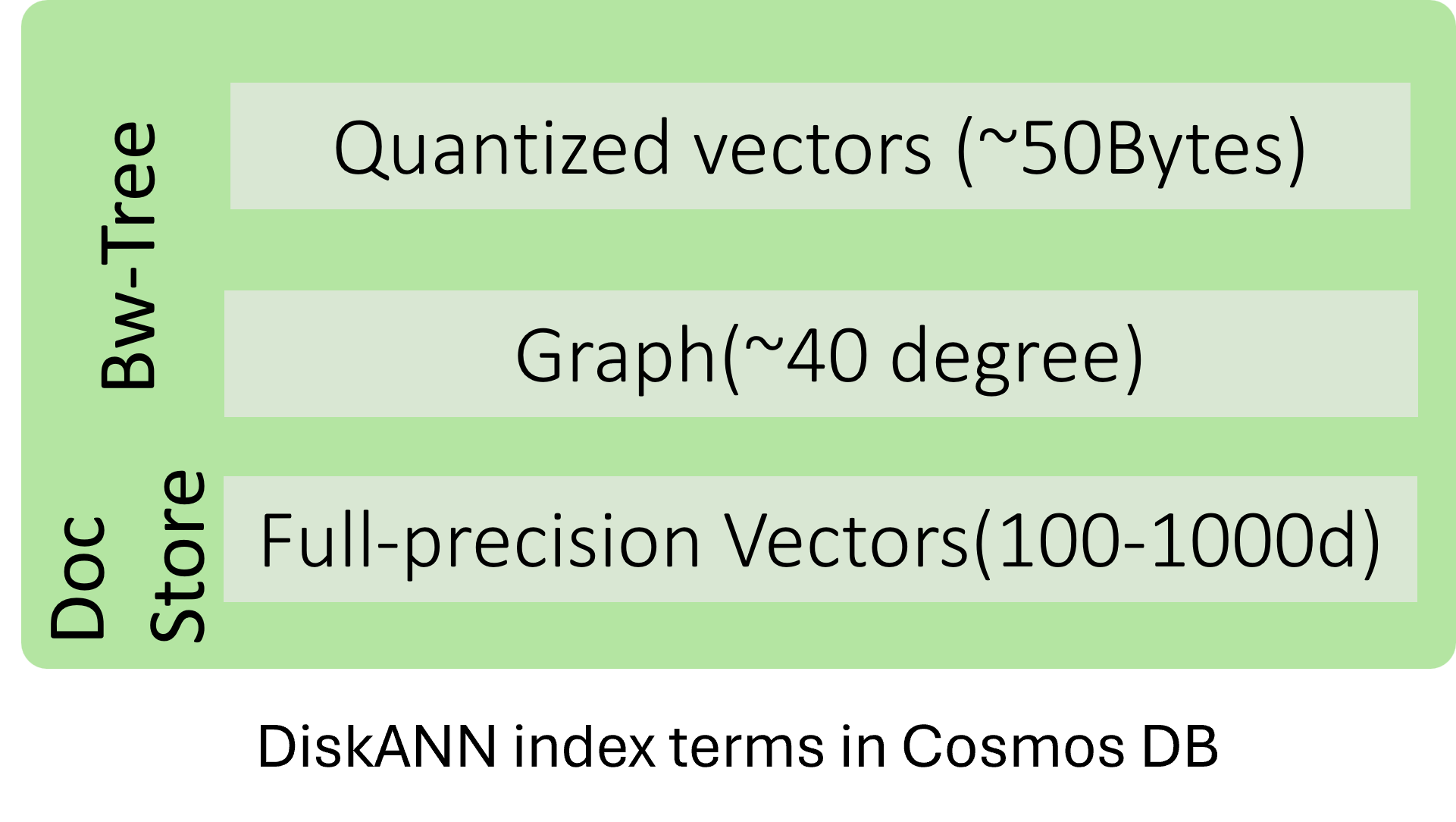}
    \caption{A comparison of the storage media for monolithic DiskANN for SSD~\cite{diskann19}
    and the newer stateless DiskANN system coupled with Cosmos DB indexing data structures.}
    \label{fig:diskann-ssd-cdb}
\end{figure}

However, an exhaustive scan in quantized space does not scale to larger replicas.
\cosmosdb can fit over 10 million 768-dimensional floating point
vectors in one replica, and we need an index that can answer queries
by accessing fewer nodes in \bwtree. We use the \diskann graph-structured
index for this. With the \diskann vector indexing policy in a collection,
two additional indexing terms in the \bwtree are created: quantized vectors (akin to Q-Flat)
and graph adjacency list terms which represent out-neighbors of each
vector in the index (see Fig.~\ref{fig:diskann-ssd-cdb}).
In the rest of this section, we  describe how we layer
these index terms in \cosmosdb, and how \diskann manipulates the terms to
update and query the index.

\subsection{Re-designing DiskANN for databases}
\diskann was previously written as a library that manages
its own buffers and index layout, thus limiting its use inside a database.
We rewrote it in Rust using the following principles 
so that it can be used with many systems including 
databases and key-value stores.

\paragraph{Decoupled index layout.}
The index layout is not controlled or even visible to the algorithms.
The core of the library consists
of methods that update and query the index by reading and updating index terms
-- quantized vectors, full precision vectors and neighbor list for vertices --
via implementations of  {\tt Provider} traits.
The {\tt NeighborProvider} trait, for example, defines the way the index would
retrieve, append and overwrite the out-neighbors of a vector.
These traits are implemented by the database which specifies the 
layout and encoding/decoding for each kind of indexing terms.
The database also owns the persistence and recovery of these terms.

\paragraph{Asynchronous interface.}
The term that the index needs may be immediately available in 
a memory buffer or may need to be retrieved from a slower storage device.
Therefore, the {\tt Provider} traits allow for {\tt get} methods 
to  return either the actual data or a future that 
eventually returns the data so that the calling thread can be scheduled with other work 
meanwhile. This is encoded via the {\tt MaybeDone} enumerated type in the snippet below:

\begin{lstlisting}[language=Rust]
/// Get the quantized vector for given `vector_id`
fn get_quant_vector(&self,
    context: &Context,
    vector_id: Data::VectorIdType,
) -> MaybeDone<impl Future<Output = 
                Result<Self::Element<'_>>> + Send>;
\end{lstlisting}

Consequently, all update and query methods in the library are also asynchronous
and need a runtime to manage threads and drive the futures to completion. 
We use the {\tt Tokio} runtime~\cite{tokioruntime}.

\paragraph{Execution Context.}
Since the library does not own any index terms, one instantiation
of the \diskann process is sufficient to update and query all the replicas
in a machine which belong to many different collections.
The database process invoking \diskann methods uses the execution
context variable to identify the target
replica for each request:
\begin{lstlisting}[language=Rust]
pub async fn insert(&self,
    context: &Context,
    vector_id: Data::VectorIdType,
    vector: &[Data::VectorDataType],
) -> Result<()>;
\end{lstlisting}

The {\tt insert} method can in turn pass this through to the {\tt Provider}
methods such as {\tt get\_quant\_vector} to help the database identify the 
term in the correct replica.
The  context can also contain LSN and {\tt activity\_id} to help
the database emit telemetry for debugging and fine-grained performance metrics.

Our rewrite achieved these goals without compromising on performance
compared to monolithic ``in-memory'' libraries such as~\cite{HNSW-github, diskann-github}.
We can implement the {\tt Provider} traits using a type backed by memory buffers
for maximum performance -- in fact, the new library is at least as fast
as the  previous monolithic \diskann library for all use cases it supported.

Further details on how inter-operation between \cosmosdb and \diskann including
runtime configuration 
and the design of C++/Rust asynchronous callbacks
are in Appendix~\ref{sec:interop}.

\subsection{Adaptations to the algorithm}
\paragraph{Querying in quantized space}
Given the  limited  memory available to the  indexer  
and query processor, Algorithms~\ref{alg:insert}
and ~\ref{alg:greedysearch}  would be too slow since full precision vectors cannot be cached
and require random reads in to the SSD.
\footnote{It is hard to re-arrange high-dimensional vectors  for spatial locality in queries, 
so each full precision vector read requires reading a random offset into the index.}
So we modify the search algorithm to traverse the graph using distance
between query and the quantized representation of the vectors which can be cached.
As observed previously~\cite{diskann19}, this does not significantly impact the convergence
rate of greedy search. However, we must re-rank a small set of best candidates found in quantized space
using distances to full-precision vectors (see Fig.~\ref{fig:rerank}).
We configure the index so that a  query for top-$10$ entries on a graph with degree $32$
and search list size $L=100$ might touch about 3500 quantized vectors,
but only about $50$ full precision vectors.

\paragraph{Indexing in quantized space }
Inserting a vector $p$ first requires querying for it as described in
Algorithm~\ref{alg:insert} to retrieve the set of visited vertices during search. 
Prior work has demonstrated that this can be done entirely in quantized space. 
The next step is to prune the visited vertex set to get the neighbor set of $p$.
Graph based indices, as described in prior work, use full precision vectors for the prune stage. We found that during prune, computations can also be done on quantized
vectors with moderate compression rates without reducing index quality.
For example, while indices over OpenAI text-3-large embeddings can be navigated
with 128 byte PQ representations, pruning accurately 
needs 256 bytes. We use this compression level for indexing and search.



\begin{figure}
    \centering
    \includegraphics[width=\linewidth]{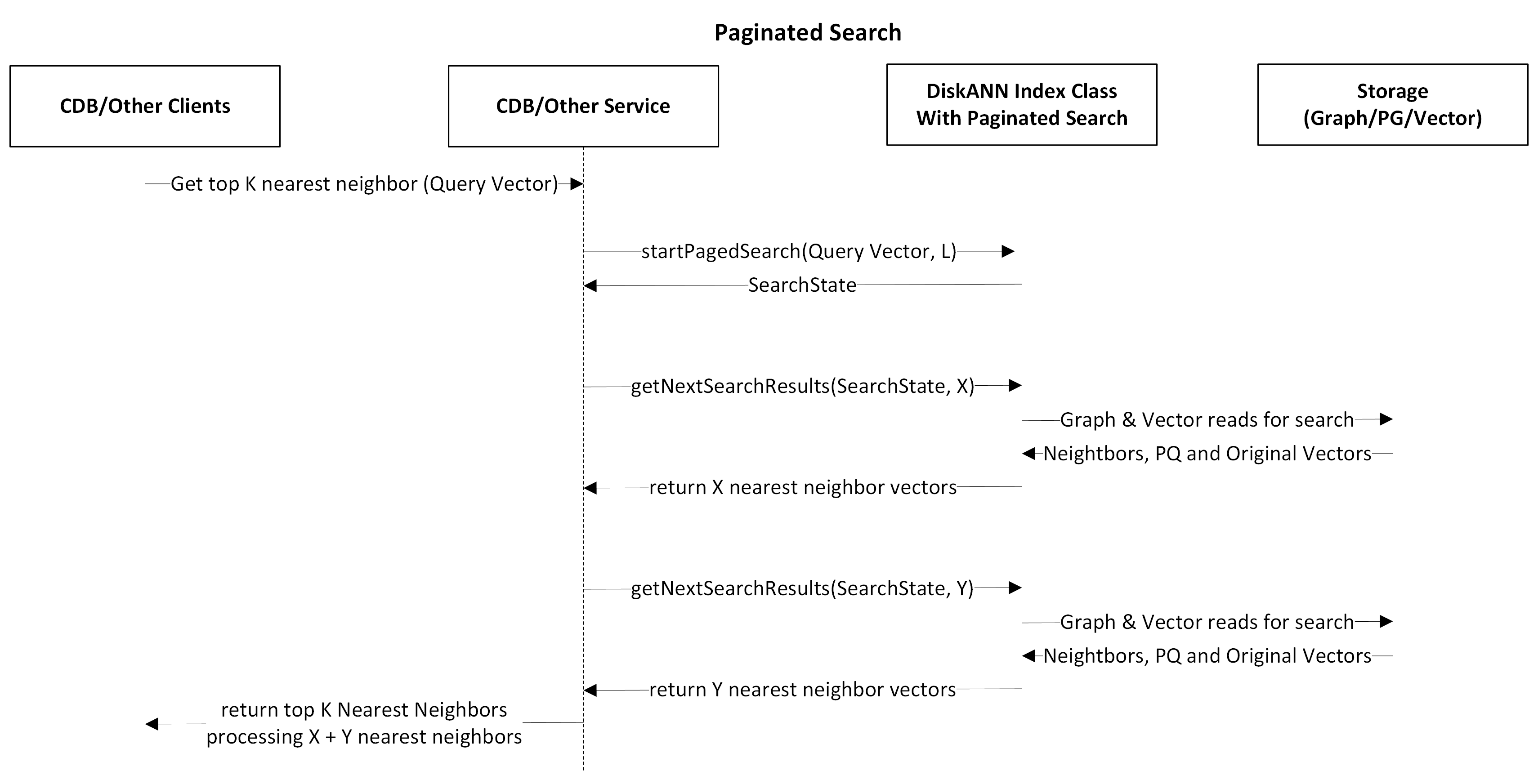}
    \caption{Control flow of paginated search.}
    \label{fig:paginated-search}
\end{figure}

\paragraph{Paginated search}
When processing hybrid queries with predicates other than similarity in vector space,
the number of candidates returned by greedy search that satisfy such predicates may be insufficient.
Therefore, we designed paginated search to allow the query layer to search
iteratively until a sufficient number of candidates that match the predicates are identified. 

Paginated search maintains two priority queues, one named {\tt best}
with max size $L$ as in the standard greedy search,
and another named {\tt backup} that has unlimited size.
Each pagination for the next $k$ candidates 
first explores the {\tt best}  queue and trims the queue to size $L$. 
Any vertices popped out from {\tt best} get pushed to the {\tt backup}  queue.
The search stops when all $L$ vertices in {\tt best}  have been visited
and returns the best $K$ results from the queue.
When the query asks for the next K, and {\tt best} does not have enough
candidates, it brings in the closest candidates from the {\tt backup} queue
and continues the search until all $L$ vertices have been visited.
A visited set saved across paginations will prevent repeated results. 
Paginated search can be performed in quantized space with appropriate reranking before returning results to the user. The control flow of paginated search is sketched in Figure~\ref{fig:paginated-search}.

\subsection{Design of index terms in \cosmosdb}
\cosmosdb stores index terms as key-value pairs in \bwtree
so they can be read via a single index key lookup or a range scan over keys. 
There are two kinds of index terms stored in \bwtree : 1) Inverted terms that map each term (path + document specific value) to a set of 'document ids' that contain them and 2) Forward terms (introduced with Vector Search) that map each term (path + document id) to any arbitrary value.

\textit{Design of Inverted and Forward Term}.
The general structure of an inverted and forward term is as follows:

\begin{itemize}[noitemsep, leftmargin=*]
\item \textit{TermKey-Prefix}: 15 bytes, the murmur hash of the property path encoded in this term.
\item \textit{TermKey-TypeMarker}: A 1 byte marker indicating the type of the value encoded in the term.
\item \textit{TermKey-EncodedValue}: A range-comparable encoding of the property's value / derived value from the user document.
\item \textit{TermValue}: An arbitrary value for the key. In practice, this is either an  Inverted Value: Variable length compressed bitmap (in buckets of 16k ranges called PES \cite{AzureDocDB}), representing the set of documents that have the property value encoded in the key OR Forward Term: Adjacency List (array of 8 byte document ids).
\end{itemize}

Next we describe the design for index terms for Quantized vectors and 
adjacency lists.
For a concrete example for each scenario, please refer to Appendix ~\ref{sec:indexterm-example}.

\begin{figure}
    \centering
    \includegraphics[width=\linewidth]{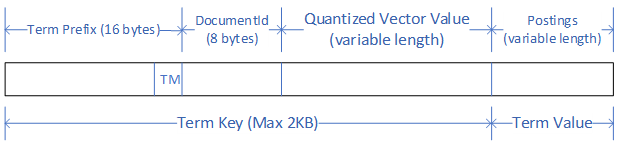}
    \includegraphics[width=\linewidth]{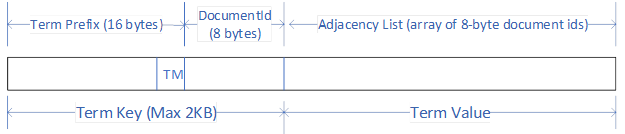}
    \caption{Index term design for quantized vectors (top) and adjacency lists (bottom).}
    \Description[Quantized Vector Design]{Quantized Vector Design}
    \label{fig:term-design}
\end{figure}

\begin{figure}[H]
    \centering
    \vspace{-10pt}
    \includegraphics[width=0.9\linewidth]{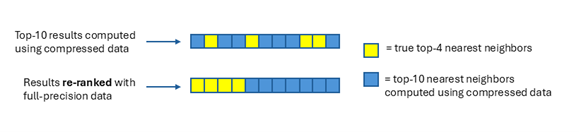}
    \vspace{-10pt}
    \caption{Re-ranking $10$ closest neighbors retrieved in quantized space 
    using full-precision vectors to find the top $k=4$. \textit{quantizedVectorListMultiplier}=2.5.}
    \label{fig:rerank}
\end{figure}

We use the  Inverted Term design to store the quantized representations of
the full vector from the user document. The \textit{TermKey-EncodedValue} includes “Document ID”
-- the 8-byte unsigned system generated unique numerical ID of the user’s document --
followed by the quantized vector in binary (see Figure ~\ref{fig:term-design}).
The \textit{TermValue} PES in this case is a dummy. 
To retrieve a quantized vector given the Document Id, a "Prefix Seek" API is used.

We designed a new ``Forward Term'' type that can have an arbitrary value rather than a bitmap (corresponding to posting lists). The 'Adjacency list terms' use the new forward term format. The term encodes out-neighbors of the graph vertex representing the vector in the document. This is done by \textit{TermKey-EncodedValue} consisting of Document Id representing the vector and the \textit{TermValue} consisting of the list of Document IDs of the out-neighbors (see Figure 4). The new value format supports blind incremental updates to the adjacency list to support fast appends. A new corresponding merge callback procedure is added to process the blind updates and consolidate on to an effective value during \bwtree consolidations. It would have been more natural to model Quantized Data with Forward Term as well,
and plan to reconcile this in the future.

\textit{Extending Term Design for sharded index}. 
By default, we construct one vector index across
all the documents in the replica. In some cases, the user might want
to query data that matches a shard key. In such a case, it is inefficient to query
the entire vector index filtering for the intended shard key. We instead
allow the user to declare a ``Sharded \diskann'' with a vector index policy
to create one \diskann index per value of the shard key present in the replica.

To support such indices, we extend the above term design by prefixing TermKey-EncodedValue with a hash of shard key value. This allows us to access both Quantized and Adjacency terms for a given shard and also co-locates the terms for a shard in a continuous key range making it easier to cache for highly active tenants. By encoding each logical shard index as just another set of \bwtree keys with a different prefix, the decoupling of logical and physical terms helps store a long tail of tenants on a single replica.

\subsection{Index construction and maintenance}\label{sec:construction}


\paragraph{\qflat Index}
The \qflat index needs a sample of vectors to create the quantization schema that is needed for generating quantized vector terms. We empirically found about 1000 samples to be sufficient for creating a first, if not the best, schema for PQ. Once the quantization schema is available, the quantized vector terms are generated inline with the document updates. A separate background process backfills the quantized terms for existing vectors. The \diskann index terms are a superset of the \qflat index and leverage quantized vector terms.





\paragraph{Graph Operations}
Updates to the graph index to reflect insertions, deletions and replications are
performed outside the transactional path to allow  \cosmosdb to maintain latency SLAs for transactions.
We calibrate the charges for transactions with vector operations to rate limit them
so that the graph index running in the background can catch up to transactions.

\paragraph{Upfront charging}
\cosmosdb charges RUs for processing vectors upfront during the transaction depending on the number of vectors and size of each vector to offer predictability to users.



\paragraph{Re-quantization}
As vectors are ingested in a collection,
we resample 25000 vectors to generate a higher quality PQ schema.
Post schema generation, quantized terms for all vectors are re-generated in place.
Newly ingested vectors are  quantized with the updated schema. 
We support distance computation between vectors quantized with two related schemas.
Since the refined schema is very similar to the original, such distance calculations are meaningful, and further,
we do not need to rebuild the graph after re-quantization. 

\subsection{Query layer}

\cosmosdb supports vector search using a built-in system function called VectorDistance
(formally described in Appendix~\ref{sec:vec-dist}).
Below is an example of a vector search query in \cosmosdb.

\begin{lstlisting}[language=SQL]
SELECT TOP 10 c.title FROM c 
WHERE c.category = 'category1'       -- Non-vector filter
ORDER BY 
VectorDistance(
    c.embedding,               -- Embedding path to match
    [0.056,-0.02,...,0.014],           -- Query embedding
    false,       -- Approximate neighbors (true is exact)
    { 'searchListSizeMultiplier':10,     -- search params
      'quantizedVectorListMultiplier':7})
\end{lstlisting}






By default, the query engine scans all documents to compute Vector Distances. When either Flat or Q-Flat index is present, the query layer instead uses them.
If quantized terms are used to compute  distance to the query,
the query engine finds 
\begin{math}
k' = \mathit{quantizedVectorListMultiplier} \times k
\end{math}
closest vectors to the query in the quantized space, and reranks them to estimate the true top  $k$.
Re-ranking is done by loading the documents corresponding to $k'$ quantized vectors from the store
and re-ordering documents based on  distance between the query and  the full precision vector.


If a \diskann index is present, then the query engine calls the Paginated search
API to get the number of documents required by the re-ranking step.
The parameter \textit{searchListSizeMultiplier} controls the quality of search -- 
higher values return more accurate results at higher latency and RU cost.
The $L$ parameter sent to paginated search set to $searchListSizeMultiplier* k$.

\paragraph{Filtering and runtime fallback}
Filters in the query are evaluated first followed by the ORDER BY.
For the query listed above, the filter c.category is evaluated first 
and translated in a compressed bitmap representing the set of documents that satisfy the filter.
Further query plan depends on the \textit{selectivity} of the filter --- that is,
how many of the points satisfy the filter.
If the selectivity is low, i.ei,  at least $5000$ documents satisfy the filter,
the query layer iteratively paginates \diskann search over quantized vector
until it finds at least \textit{quantizedVectorListMultiplier}$\times k$
documents that satisfy the predicate. Predicate matches
 can be checked quickly using the compressed bitmap.

\paragraph{Filter Aware Graph Search}
Given fast access to the bitmap over documents that satisfy the filter, 
we can modify \diskann search  to make it more likely to find points that satisfy the filter.
This is done by Algorithm~\ref{alg:betasearch} in Appendix~\ref{sec:interop}, which scales down the
distances from the query to vectors that satisfy the filter by $\beta<1.0$ in greedy search iteration. 
With this method, we can support the broad set of predicates that Cosmos DB supports with  filter aware graph search.

\paragraph{Highly Selective Queries}  Queries with predicates that match fewer than 5000 documents 
 are routed by the query planner to use either the quantized flat index or full precision vectors in the document.
 Query planned  uses the quantized vectors stored in Bw-tree which supports fast range scan
 when the document matching the filters are in a relatively small range.
  For queries that have match few documents scatter across large range, query planner
  decides to  load individual points from document store instead of doing a large range quant vector scan.





\paragraph{Sharded DiskANN}For larger indices with suitable distributions on their labels,
 we require a \textit{sharded \diskann index policy} to provide shard key base on label. This allows us to create \diskann instance base on the shard key and a filtered query to route directly to the shard corresponding to its label, or one of its labels.
We can compose sharded and filter-aware \diskann queries. For a multi-tenant collection configured
with sharded vector indexing, the queries with filter over the shard key as well additional filters $f$
can limit search to the relevant shard and use the beta-biased greedy search to optimize for filters $f$.

\paragraph{Continuations}
\cosmosdb backend requests are limited to 5 seconds.
If a query does not complete by this time, 
it is preempted with a continuation token
capturing the query state that the client can  use to resume the query.
While some queries such as regular ORDER BY queries are streaming, pagination in quantized space is not.
The partial results in a vector search cannot be serialized to the continuation token as it can be large in size.
So, to handle continuation, the partial results are returned to client, which is responsible for 
reordering and merging them across continuations. 

\paragraph{SDK Query Plan:}
 \cosmosdb SDK has a query planner that can distinguish between single partition queries which are passed through and queries requiring cross partition fan out.
 The SDK supports fan out and aggregates the results from different partitions.
 It also handles continuations for non-streaming queries like Vector Distance Order By,
 and supports  collation of partial results, merging cross-partition replies and re-ranking for the final results.

\section{Evaluation}\label{sec:eval}

We now measure the query latency, cost, and ingest performance of 
our design. Our experiments scale up to 10 million vectors per partition
and scale out to 1 billion vectors across a collection.



\paragraph{Datasets} We use the following datasets in our experiments.

\begin{itemize}[noitemsep, leftmargin=*]
\item \textbf{Wiki-Cohere}: 35 million Wikipedia articles embedded using the cohere.ai multilingual 22-12 model~\cite{wikipedia}, and a query set of 5000 embeddings of Wikipedia Simple articles, with 768 floating-point dimensions.
We use 100K, 1M and 10M prefixes of this collection.
\item \textbf{MSTuring}: 1 billion Bing queries embedded to 100 dimensional
floating point using the Turing AGI v5 model~\cite{msturing}, with 100,000 queries from the same distribution.
\item \textbf{YFCC Dataset}: 1 million vectors corresponding to a 192 dimensions CLIP embedding applied to (copyleft) photos and videos in Flickr from the year 2004 until early 2014. This also includes metadata
such as camera model, country, year, month. 
The number of documents per year ranges from ~30,000 to ~144,000. 
\end{itemize}


\textit{Runbooks} are long-running sequences of insertion, deletion, and queries
to simulate various streaming scenarios~\cite{bigannneurips23, simhadri2024resultsbigannneurips23}. In this work, our first runbook is based on an expiration time model - where each point is inserted with a randomly selected expiration time, at which point it is deleted - as they are a good proxy for common production scenarios.
For expiration time model, we use two instances based on the Wikipedia-10M dataset and the MSTuring-1M dataset to benchmark the recall stability of the vector index. Our second runbook is more adversarial and meant to imitate distribution shift. This runbook instance is based on the MSTuring-10M dataset. The dataset is partitioned into 32 clusters, and points are inserted and deleted in clustered order. For each runbook, the same query dataset is used at each query step.

%

\paragraph{Configuration}
We use the the following parameters unless noted otherwise.
The graph degree is 32 to minimize footprint of the index,
and a slack factor of 1.3 is used to reduce number of secondary prunes.
We use $L=100$ for index construction.
The  parallelism for mini-batch inserts is set to 8, since
replicas might not regularly get more than 3 cores on shared machines.
Each physical partition size limit is 50GB and Bw-tree max chain length is set to 15.
When measuring query performance, the index is queried
in a warm up phase, following by 5000 queries issued one at a time.
The Bw-Tree cache is  configured to be large enough to cache the quantized and adjacency list index terms.

\begin{figure}
    \centering
    \includegraphics[width=0.9\linewidth]{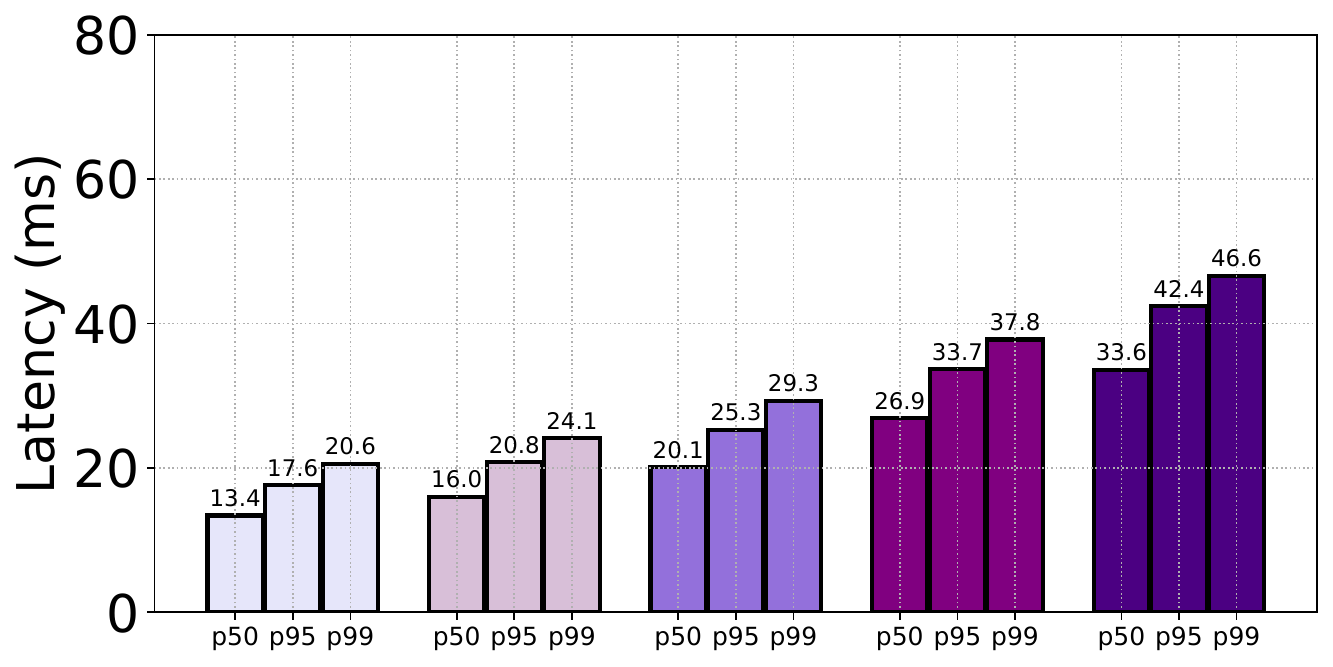}
    \includegraphics[width=0.9\linewidth]{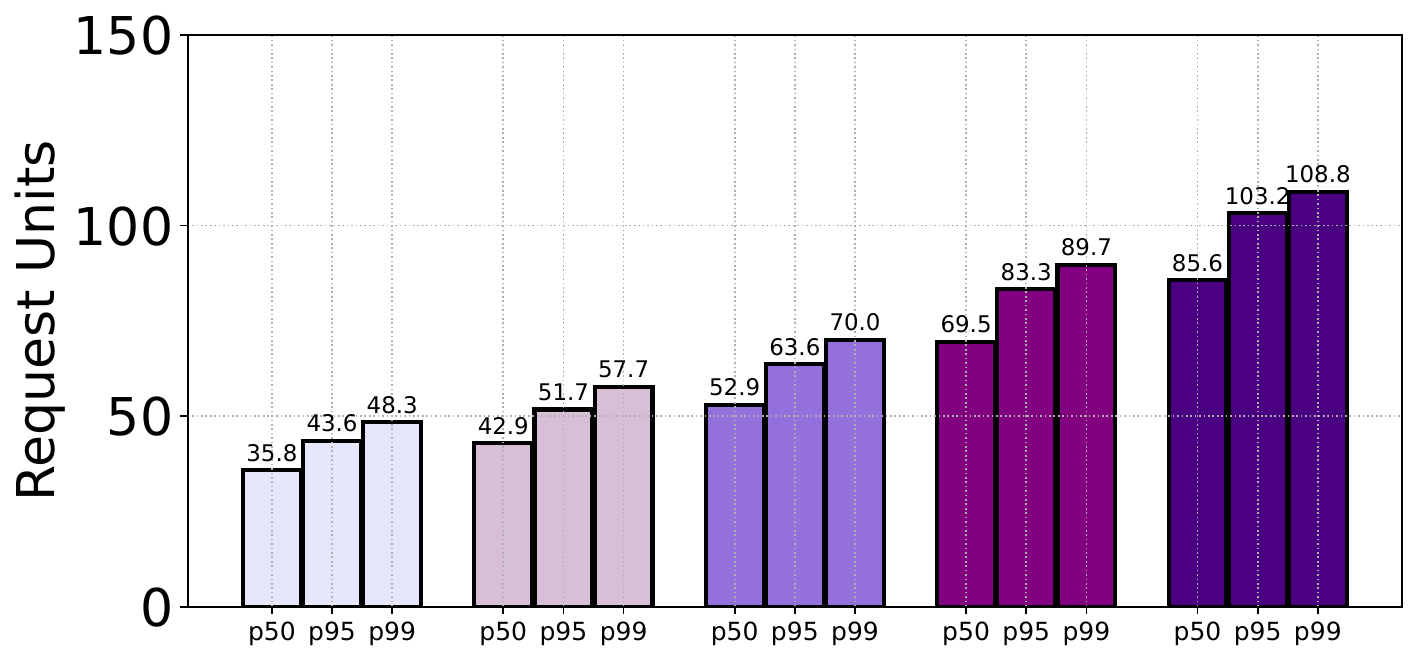}
    \includegraphics[width=\linewidth]{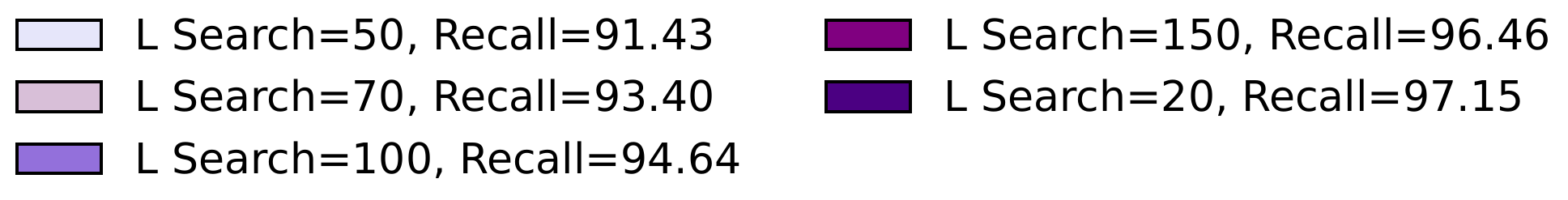}
    \caption{p50, p95 and p99 query latencies and query RU charge for 10 million Wiki-Cohere vector index for various values
    of search list size, and the corresponding recall@10. \vspace{-10pt}}
    \label{fig:wiki-10M-query-latency}
\end{figure}

\subsection{Query latency and RU charges}

Figure~\ref{fig:wiki-10M-query-latency} shows the query latency and RU charge for an index of 10~million Wiki-Cohere vectors.
Note that with LSearch=50, p50 latency is under 20~ms for the 10~million index with a recall@10 of 91.83\%.
Increasing $L$-search gives higher recall, but with increased latency and RU charges.
The corresponding plots for 1~million and 100,000 vector indices are shown in Figures~\ref{fig:wiki-1M-query-latency} and~\ref{fig:wiki-100K-query-latency} in Appendix~\ref{sec:apdx-query-plots}. 

\begin{figure}
    \centering
    \includegraphics[width=0.49\linewidth]{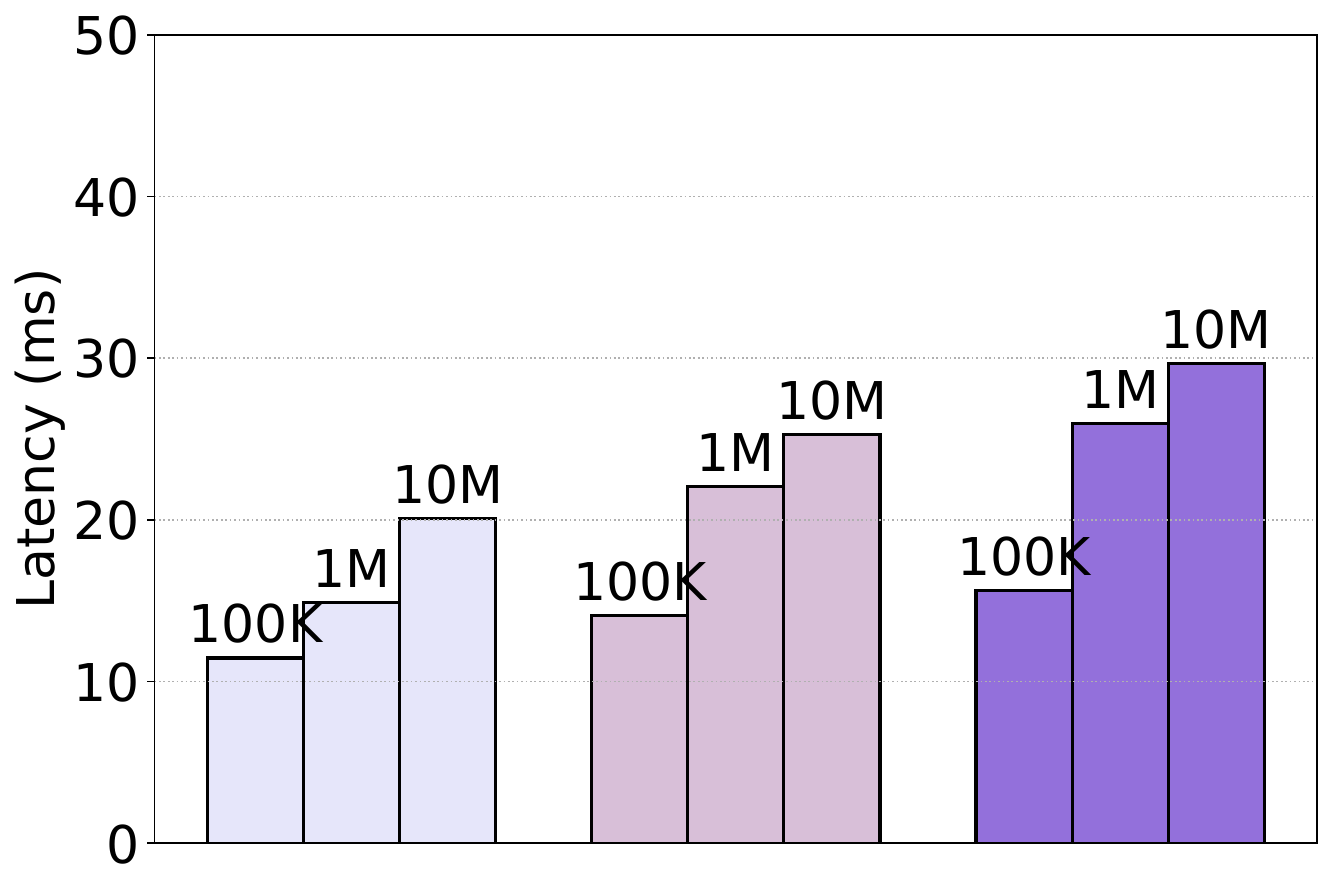}
    \includegraphics[width=0.49\linewidth]{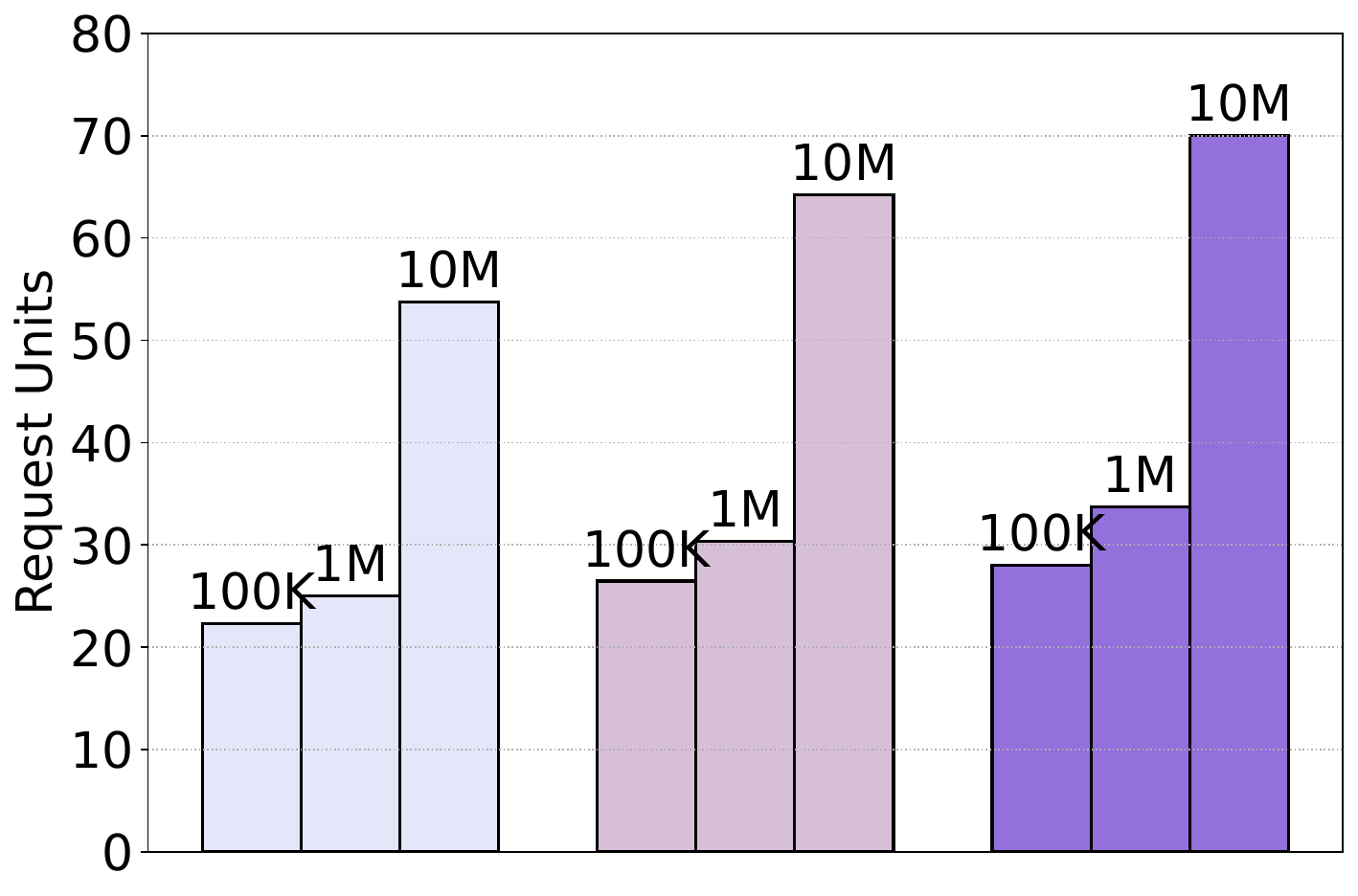}
    \includegraphics[width=.6\linewidth]{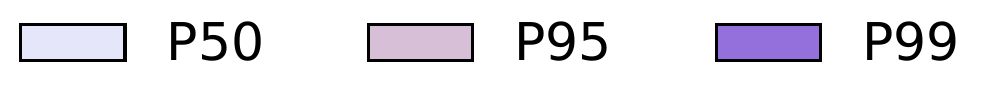}
    \caption{p50, p95 and p99 query latencies and query RU charges for  Wiki-Cohere vector indices over
    100K, 1 million and 10 million vectors with $L=100$ which provides 94.64\% recall@10.}
    \label{fig:wiki-query-combined}
\end{figure}

\begin{figure}
 \vspace{-3pt}
    \centering
    \includegraphics[width=0.49\linewidth]{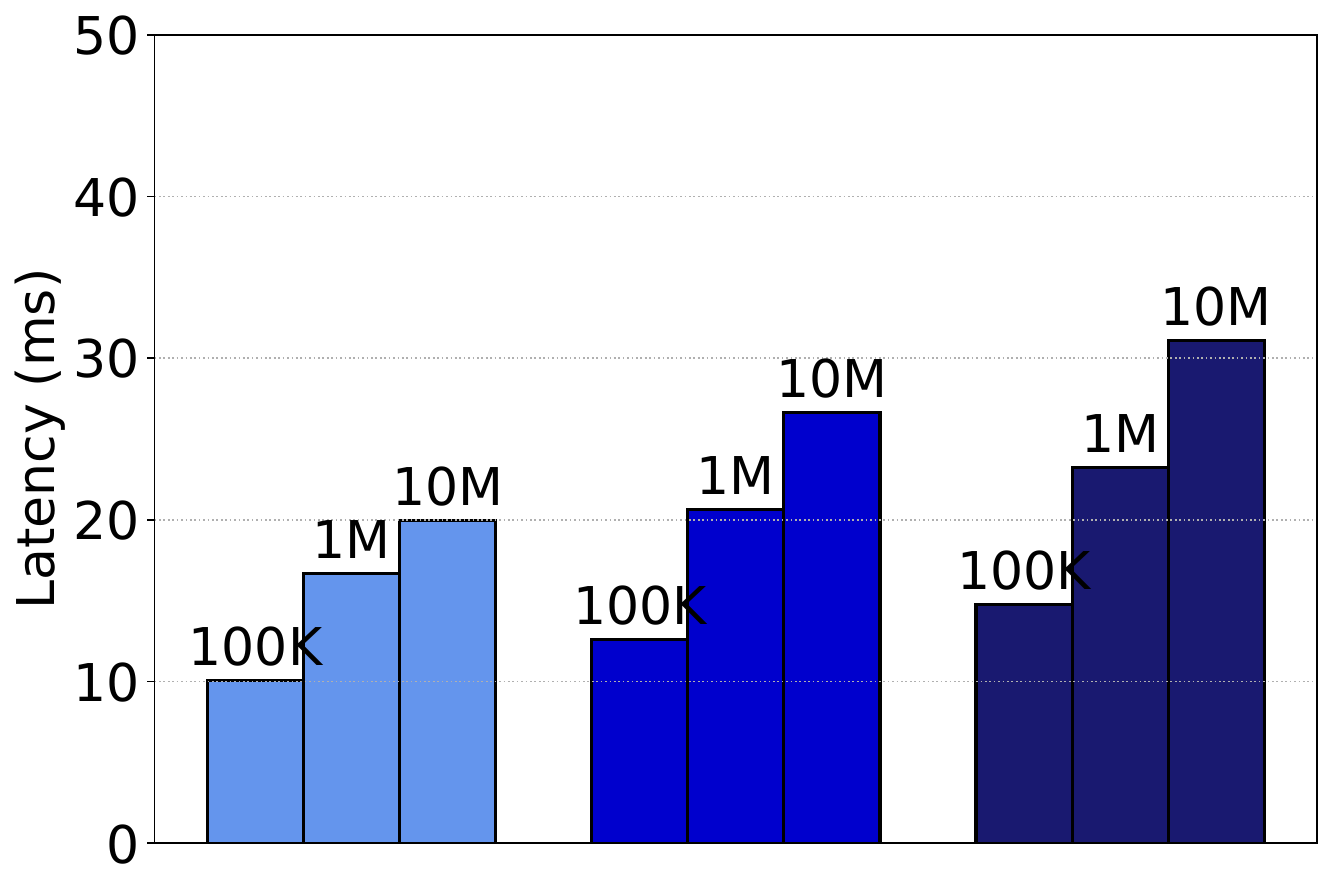}
    \includegraphics[width=0.49\linewidth]{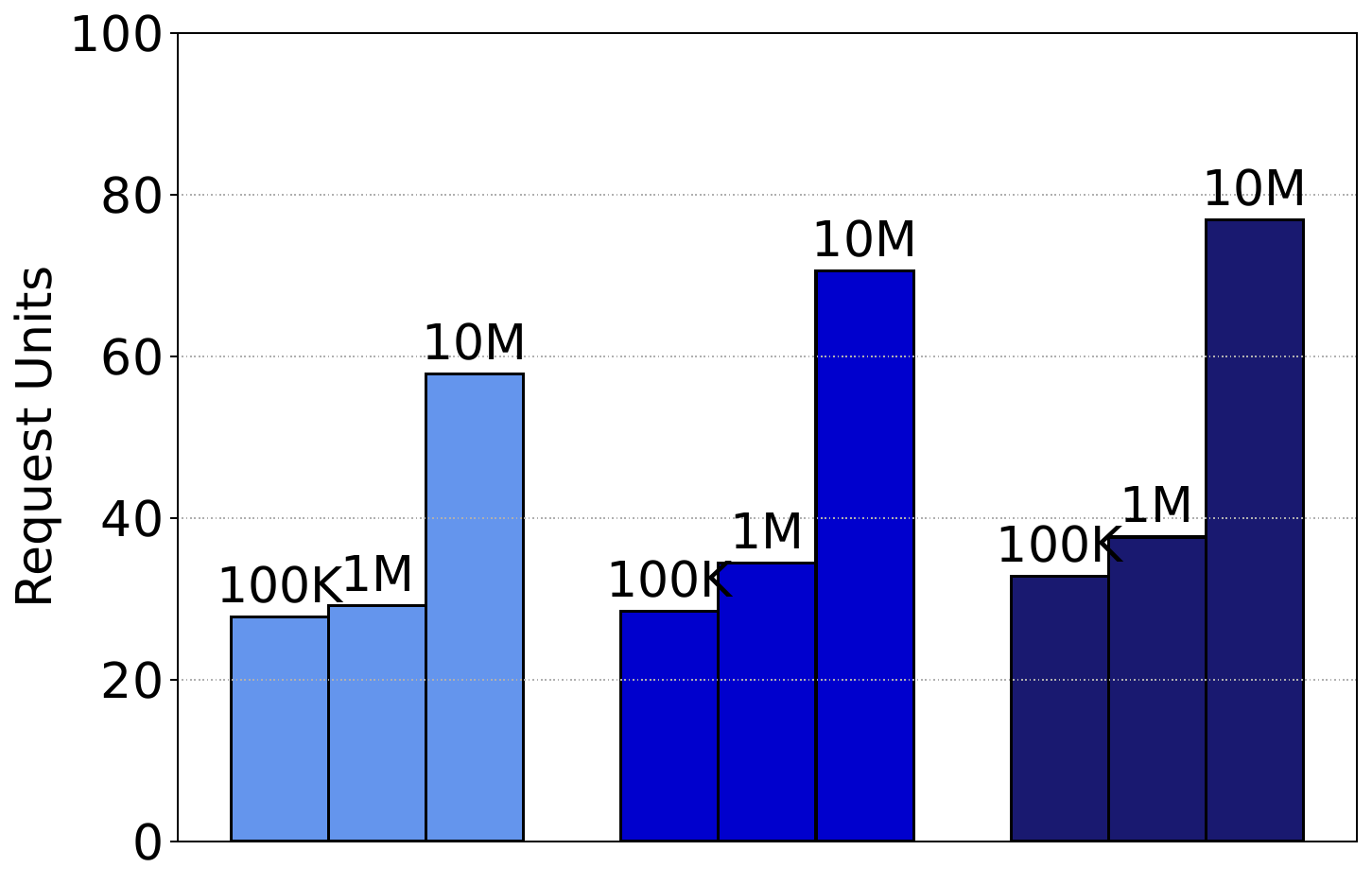}
    \includegraphics[width=.6\linewidth]{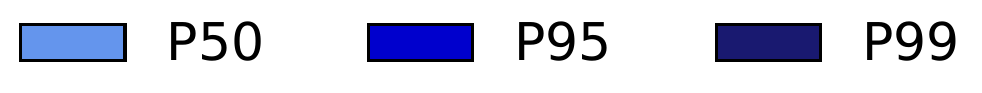}
    \caption{p50, p95 and p99 query latencies and query RU charge for MS Turing vector indices over
    100K, 1 million and 10 million vectors with $L=100$.}
    \label{fig:msturing-query-combined}
\end{figure}

We hold the search list constant and compare the query complexity as we increase
the size of the index from $100K$ to 1 million to 10 million vectors
in Figures~\ref{fig:wiki-query-combined} and~\ref{fig:msturing-query-combined}.
We note that the p50, p95 and p99 {\bf latencies increase by
less than $2\times$ despite the $100\times$ increase in index size}.
The RU charge similarly {\bf increases less than $2\times$} except in the case of Wiki-Cohere 10M.
Here, we chose a 2 partition set up (we could have also fit it in one partition).
Therefore, we pay the extra cost for fanning out to an additional partition.
Another important point to note is that {\bf despite the increase in dimensionality from 100 to 768, there is barely 
any increase in query latency or cost}. The Cosmos DB design is well suited for extremely high dimensional vectors.

We compare Azure Cosmos DB vector search with  enterprise-grade serverless vector databases --
Pinecone\cite{pinecone-enterprise-pricing}, Zilliz\cite{zilliz-enterprise-pricing}, and DataStax\cite{datastax-pricing} --
using their publicly available pricing and documentation. For Azure Cosmos DB, we use query costs determined from experimental
tests using default settings. Costs for these other services are reflect in their own cost units (Read Units, vCUs, etc.).

In Table~\ref{tab:cost-comp}, we see that for the 10  million Wiki-cohere index,
{\bf Cosmos DB has nearly $43\times$, $12\times$, and $1.75\times$ lower query cost than Pinecone (enterprise tier), Zilliz (enterprise tier), and DataStax (standard tier) respectively.}

\begin{table}[]
    \caption{P99 vector search query and monthly storage costs of enterprise-grade serverless vector DBs
    for 1 million queries over 10 million 768-dimensional vectors as of July 14, 2025. Cosmos DB provides 94.64\% recall@10. }
    \vspace{-7pt}
    \label{tab:cost-comp}
    \centering
    \begin{tabular}{rcccc}
    \toprule
           &  RU per  & \$ per & \$ per  & Storage  \\ 
           & Query  & 1M RUs & 1M queries & Cost (\$) \\ \hline
        Cosmos DB  & 70                             & \$0.25        & \$17.50  & \$22.25 \\
        Pinecone~\cite{pinecone-enterprise-pricing} & 32            & \$24   & \$768 &  \$11.55 \\
        Zilliz~\cite{zilliz-enterprise-pricing}     &  55           & \$4    & \$220 & \$17.84 \\
        DataStax\cite{datastax-pricing} & 768 & \$0.04 & \$30.72 &  \$24 \\

\bottomrule
    \end{tabular}
    \vspace{-15pt}
\end{table}

 \begin{figure}
 \centering
 \includegraphics[width=0.49\linewidth]{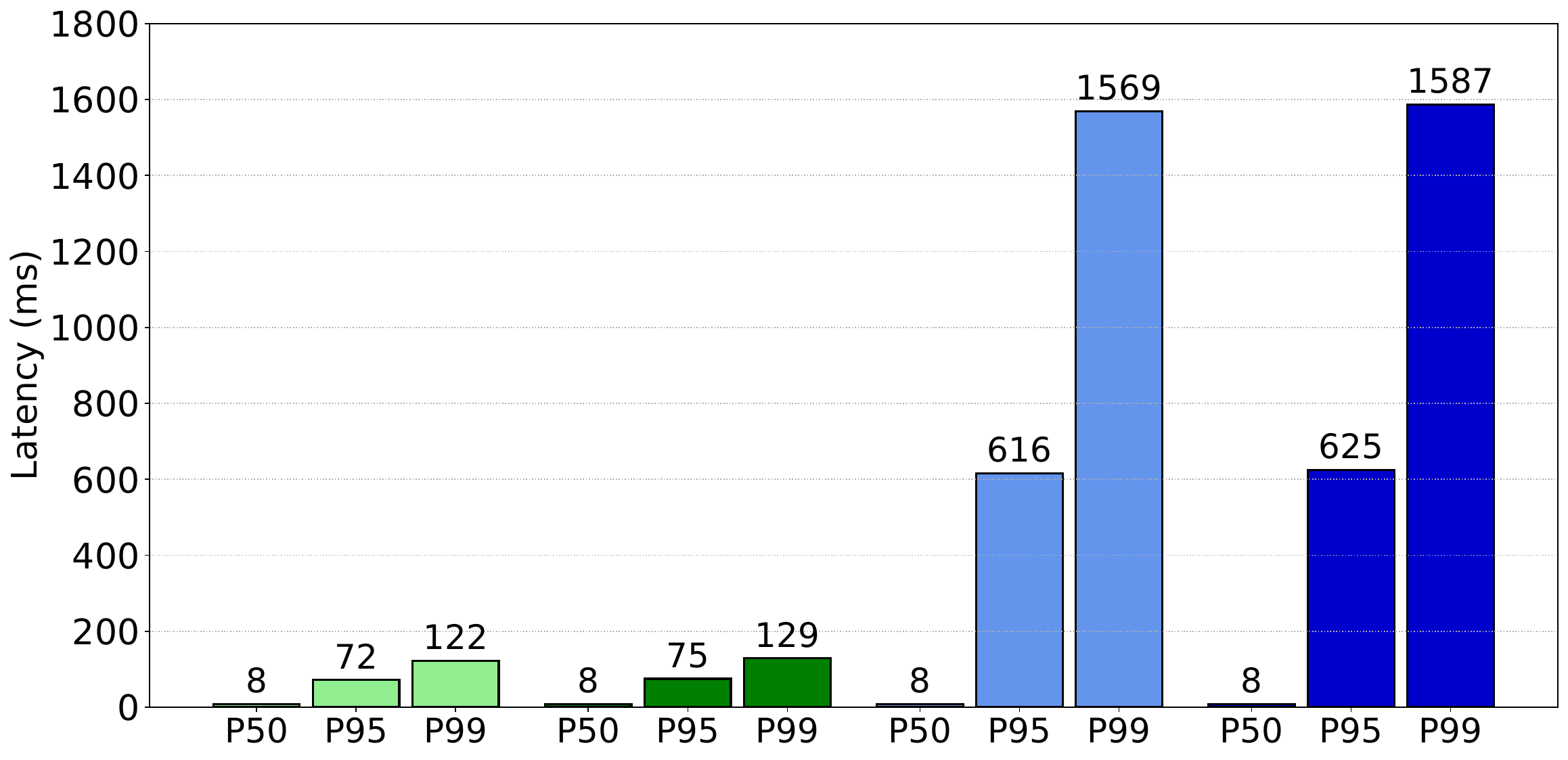}
 \includegraphics[width=0.49\linewidth]{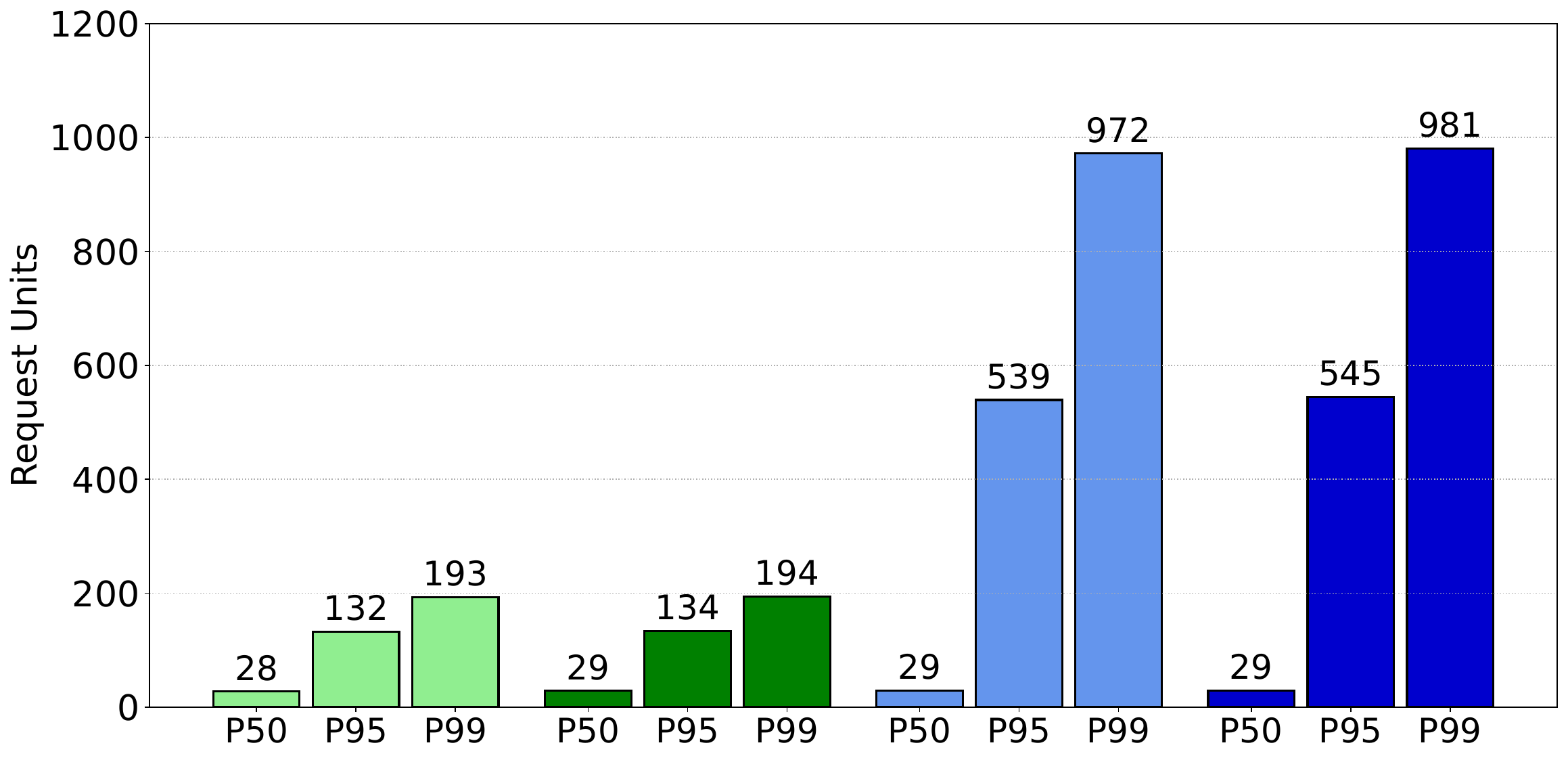}
 \includegraphics[width=\linewidth]{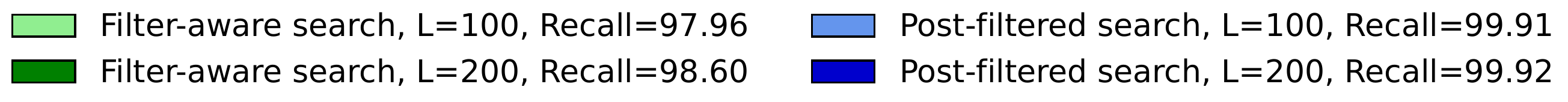} \\
 \caption{Comparison of the query latency and cost of post-filtering and filter-aware search on the YFCC dataset.}
 \label{fig:filtered}
 \end{figure}

\begin{figure}
    \centering
    \includegraphics[width=0.49\linewidth]{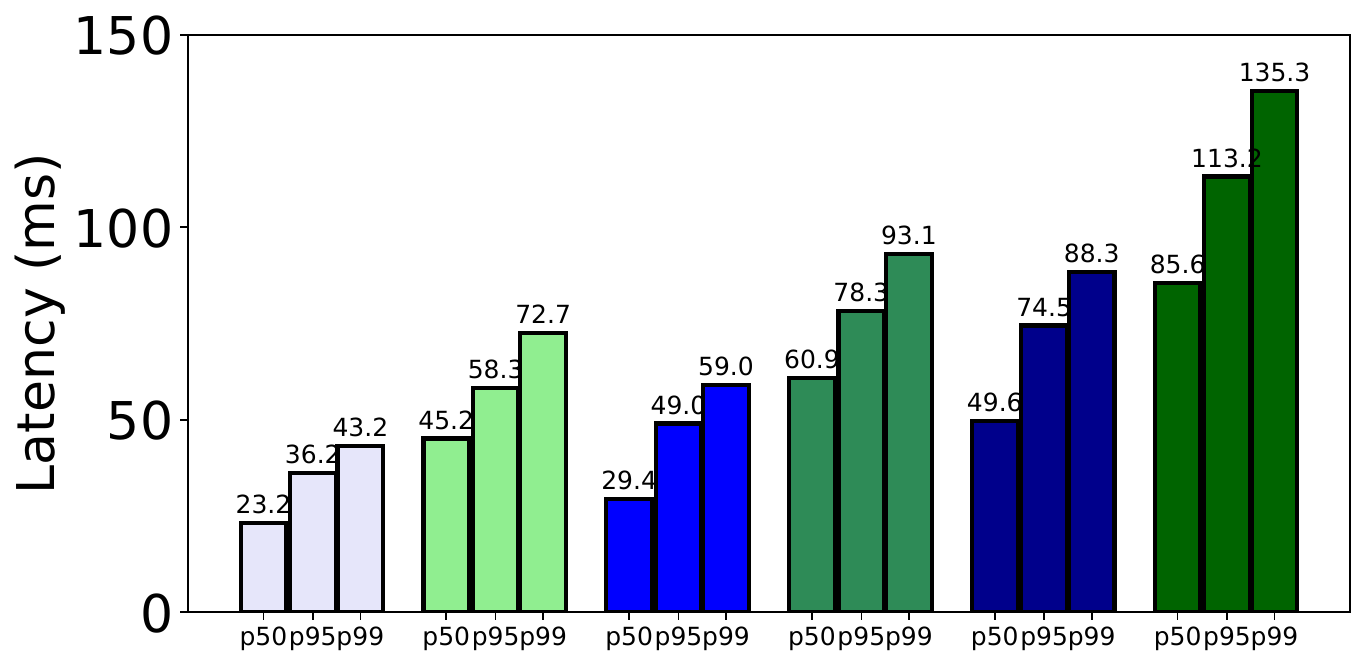}
        \includegraphics[width=0.49\linewidth]{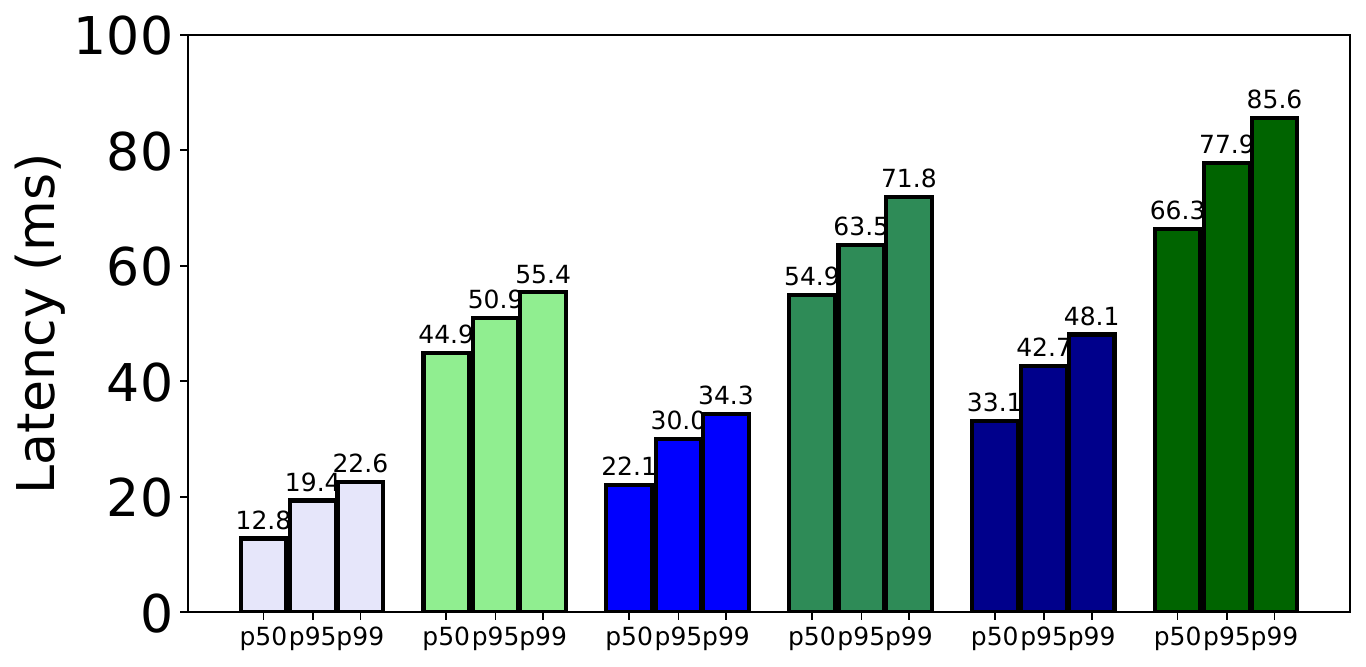}
    \includegraphics[width=\linewidth]{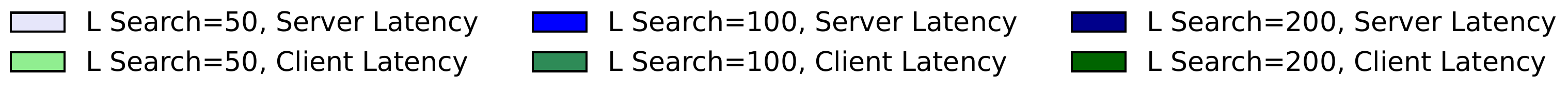} 
        \includegraphics[width=0.49\linewidth]{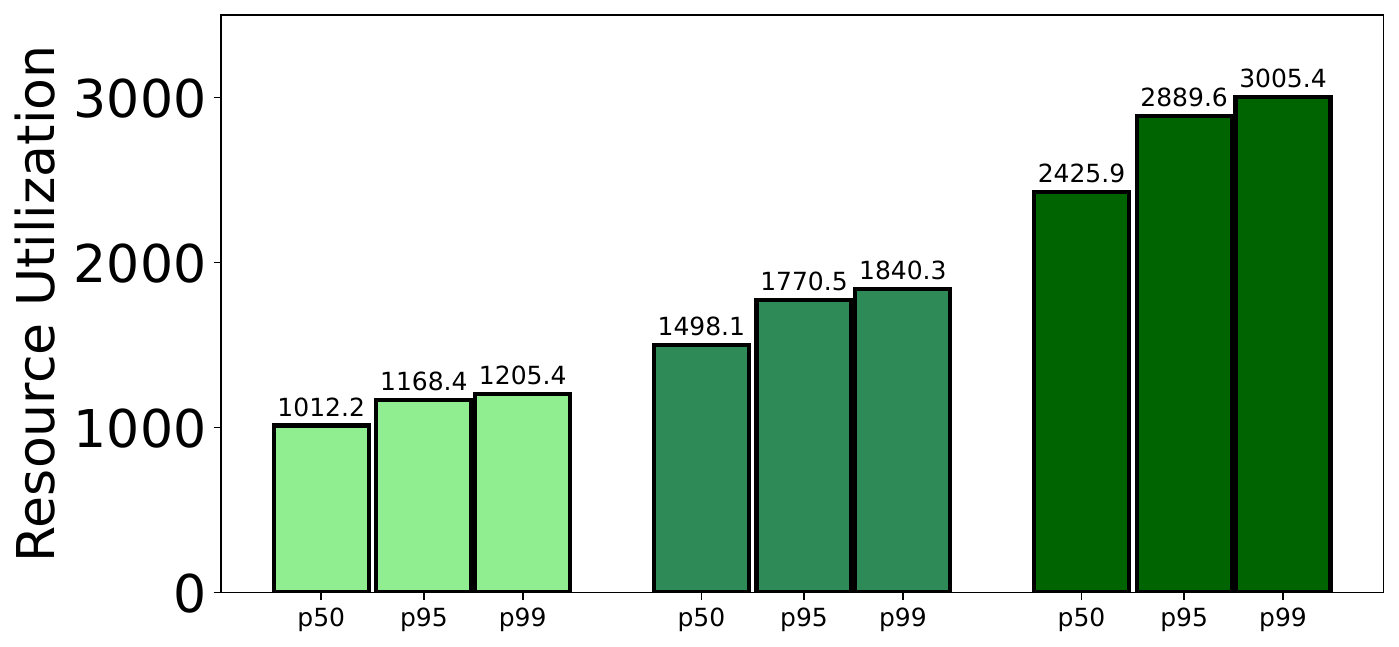}
        \includegraphics[width=0.49\linewidth]{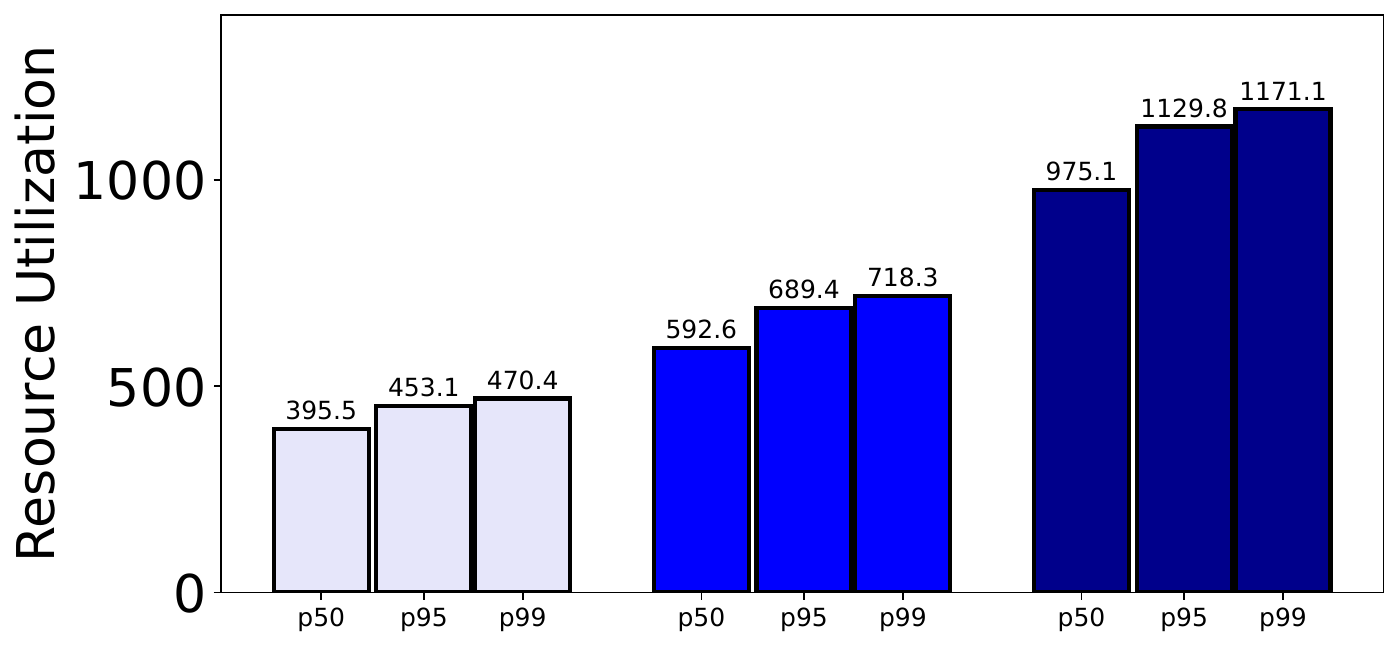}
    \includegraphics[width=\linewidth]
    {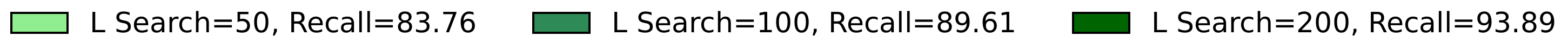}
    \includegraphics[width=\linewidth]
    {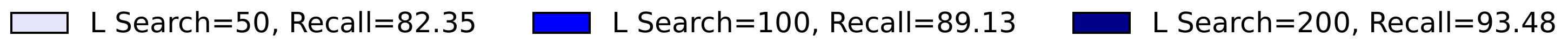}
    \caption{
    Query latency and RU charge for 1 billion (Left) and 100 million (Right) MS Turing vector index collections. }
    \label{fig:msturing-1B}
\end{figure}

 \subsection{Filtered queries}
 Next, we consider the recall, latency, and the cost of two approaches for filtered queries using
 the YFCC dataset in Figure~\ref{fig:filtered}. We compare the post-filtering search
 -- an unmodified graph search which filters the final answer to return only points satisfying the filter --
 and filter-aware (Algorithm~\ref{alg:betasearch}, $\beta=0.3$) approaches.
 
 When the number of database points matching the filter is about 5000, the query planner
 sends queries invokes the \diskann index with either post-filtering or filter-aware search.
 Otherwise, query planner falls back  to simpler look ups as described in Section\ref{sec:design}.
 
 We find that both approaches are capable of providing high recall at low latency and RU cost.
Post-filtering search produces slightly higher recall than filter-aware search for the same $L$ value,
 but it suffers from diminishing returns in recall and high p95 and p99 latency.
 On the other hand, filter-aware search's recall can be improved by increasing the search depth ($L$),
 while still retaining predictable tail latency and RU cost.
 
 For instance, $L=200$ with filter-aware search yields the similar recall as $L=200$ with post-filtering
 search, but with 10x better p99 latency and 5x less p99 cost.
 This suggests that filter-aware is better for controlling resource utilization and tail latency.

\subsection{Scale out and Query Aggregation}
Cosmos DB can scale up to millions of vectors per partition (subject to the 50 GB limit on data and index),
and  scale out to  billion-scale vector indices.
Figure~\ref{fig:msturing-1B} shows the query latency and RU properties of such collections.
The MS Turing 1 billion vector collection spans 50 partitions, while the 100 million vector collection spans 
20. The server latency is measured per partition
and client latency is the end to end latency observed in client side.
We use a  E64is\_v3 Azure VM with 64 vCPUs on the same Azure region as the index
to aggregate query results with  maximum concurrency and minimum latency.
Since the query fans out to large partitions concurrently, 
the client end-to-end latency is sensitive to the worst latency on the server side. 
The best latency is achieved by using fewer partitions packed with as many vectors as possible.
The same applies to RU charges since the query cost grows logarithmically with the number of
vectors in a partition, but linearly with the number of partitions.

\subsection{Ingestion}
We now study the vector ingestion component that updates the Bw-tree with index terms reflecting new vector insertions.
The parallelism for batch inserts is set to 32 threads, The underlying machine has 16 physical and 32 vCPUs.
We set the cache to be large enough to hold the vector index terms in memory (excluding full precision vectors).
We set the graph degree $R=32$, index build parameter $L_{build}=100$ and \bwtree chain length set to 15. 
Each insert typically accesses approximately $R \cdot L_{build}$ quantized vectors and just over $L_{build}$ adjacency lists (with a wide margin of error).

\begin{figure}
    \centering
    \includegraphics[width=\linewidth]{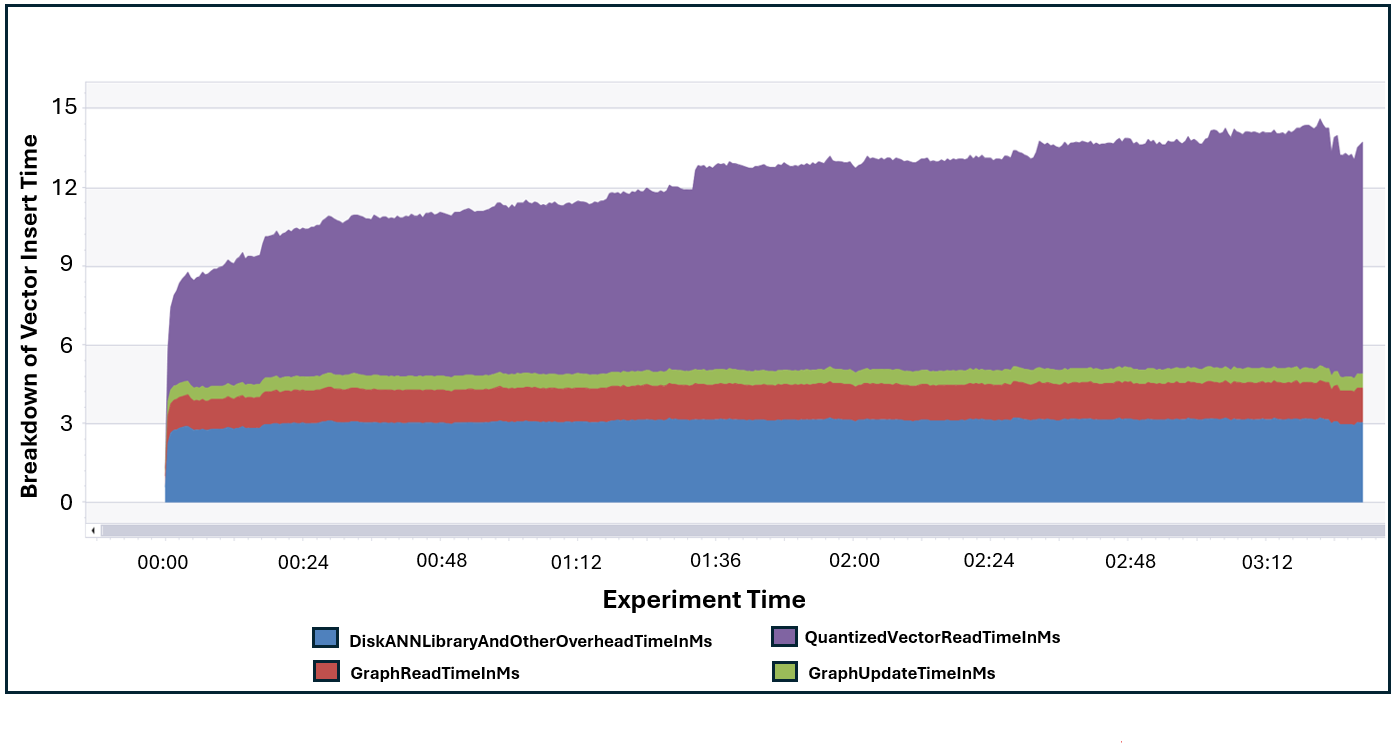}
    \caption{Latency breakdown for single threaded MS Turing-1M ingestion with components (Quantized vectors, Adjacency list, DiskANN library), Bw-Tree chain length = 15.}
    \label{fig:ingest-breakdown}
\end{figure}

Figure~\ref{fig:ingest-breakdown} provides a breakdown of the CPU time spent in each insert.
Most of the time is spent accessing the quantized vectors, as expected.
Since they are cached, this operation is synchronous and involves
traversal of cached \bwtree terms. The next two dominant components are the time
spent in DiskANN library and reading the adjacency lists. The actual distance computations as well
as updating candidate sets in \diskann is small -- under 3ms/insert.

\begin{figure}
    \centering
    \includegraphics[width=\linewidth]{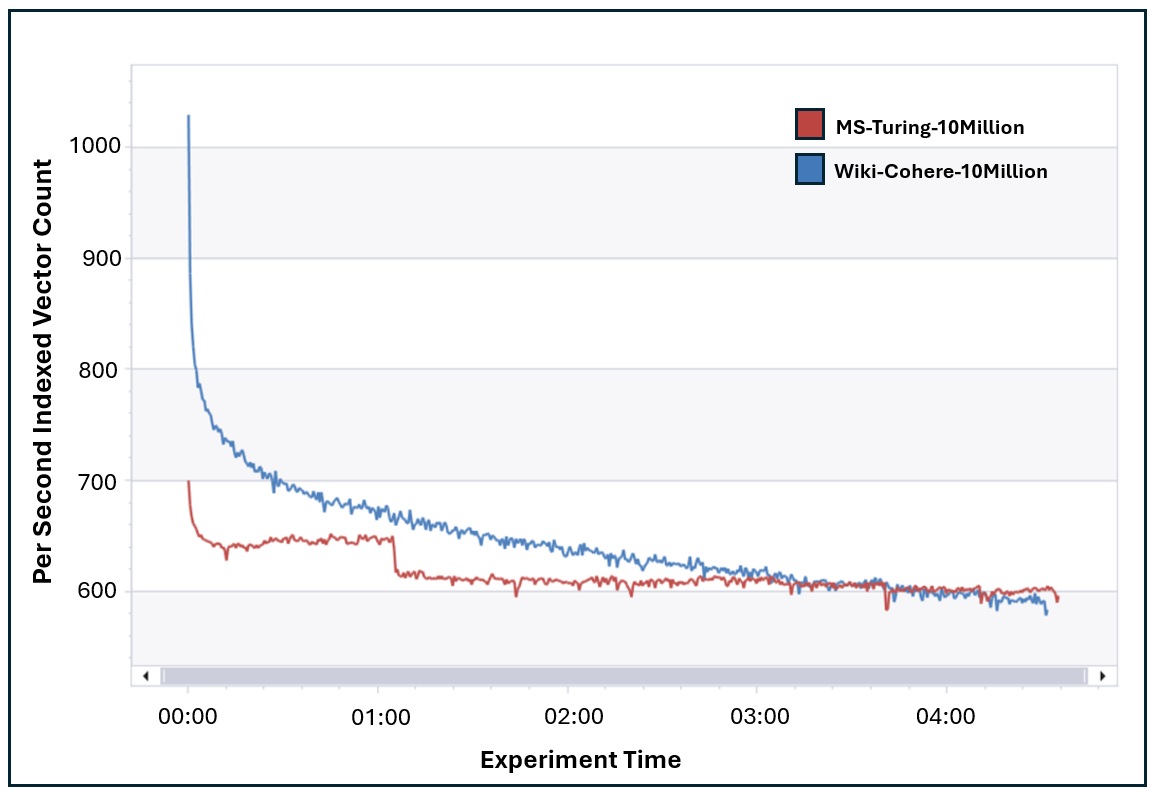}
    \caption{Ingestion progress on MS Turing-10M and WikiCohere-10M datasets when
 Bw-Tree cache is large enough for vector index terms.}
    \label{fig:ingest-progress}
\end{figure}

Figure ~\ref{fig:ingest-progress} shows ingestion progress with Resource Governance disabled 
to make it easier to analyze data. In practice, Resource Governance controls compute available to a single replica
to ensure fairness. Earlier in the ingestion, the index cost to access quantized vector terms is lower,
and the rate of ingestion is higher. The ingestion rate gradually decreases as ingestion progresses since Bw-tree chain
lengths grow longer and each quantized vector lookup takes more microseconds.
The estimated CPU time needed for a single single vector insert  -- with $R = 32$, $L_{build} = 100$,
and $10\mu s$ for single quantized vector read, and $25\mu s$ for  graph adjacency list read,
and $3ms$ spent in DiskANN library -- is about 25ms.
This yields a throughput of ~40 vectors/sec per thread, or ~640 vectors/sec on a 16-core machine, 
which matches the steady state measurement in Figure~\ref{fig:ingest-progress}

We also evaluate vector ingestion costs on the WikiCohere-10M dataset, using experimental data for Azure Cosmos DB and compare it with the published pricing and cost guidance for the enterprise-grade serverless vector databases their respective costing units equivalent to RUs (i.e., Read Units for Pinecone~\cite{pinecone-enterprise-pricing} and DataStax~\cite{datastax-pricing} and vCUs for Zilliz~\cite{zilliz-enterprise-pricing}) in Table~\ref{tab:insert-cost-comp}.  Cosmos DB demonstrates lower vector insertion costs—approximately 33\% and 53\% less than Pinecone and DataStax (standard tier), respectively.

Azure Cosmos DB does have $5.4\times$ higher insertion charges than Zilliz. However, provisioned throughput and autoscale billing models offer at least a $7\times$ price discount on RU charges compared to serverless, so these higher costs can be further reduced.

\begin{table}[]
    \caption{Insertion costs of enterprise-grade serverless vector databases
    for 10 million 768D vectors (Wiki-Cohere-10M) as of July 14th, 2025.}
    \vspace{-10pt}
    \label{tab:insert-cost-comp}
    \centering
    \begin{tabular}{rcccc}
    \toprule
        & \$/1M RU & RU/insert & Total cost (\$) \\ \hline
        Azure Cosmos DB    & \$0.25        & 65 & \$162.5 \\
        Pinecone~\cite{pinecone-enterprise-pricing}  & \$6   & 4 &  \$240 \\
        Zilliz~\cite{zilliz-enterprise-pricing}        & \$4  & 0.75 & \$30\\
        DataStax\cite{datastax-pricing} & \$0.04 & 768 & \$307.2 \\
\bottomrule
    \end{tabular}
\end{table}

\subsection{Recall Stability over Updates}

\begin{figure*}
\centering
\begin{subfigure}{.33\textwidth}
    \includegraphics[scale=.33]{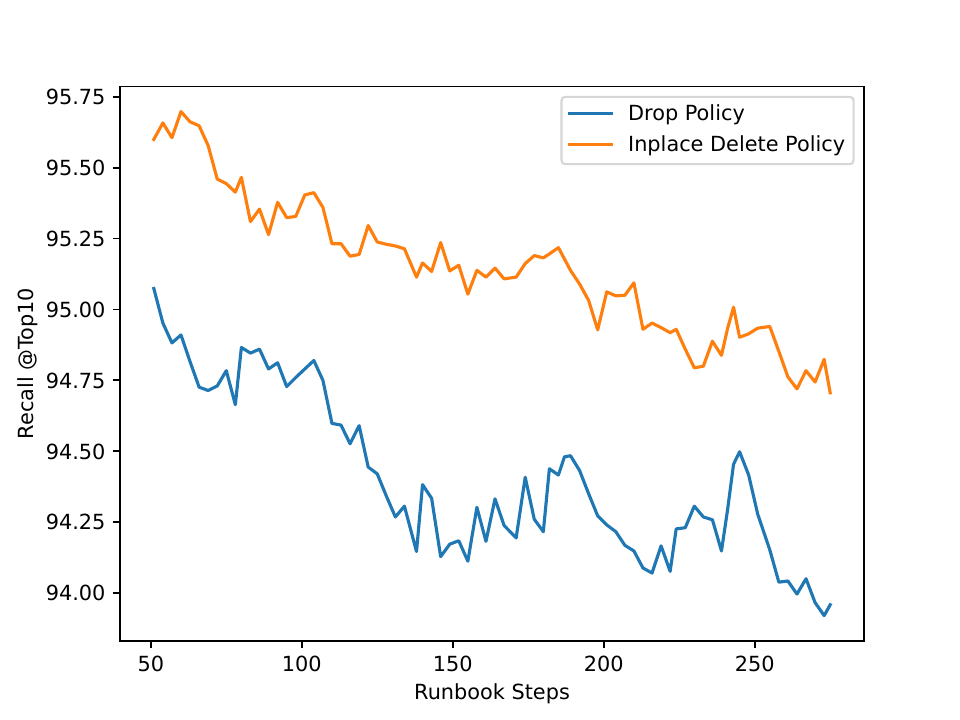}
    \caption{WikiCohere-10M expiration time runbook.  }
    \label{fig:wiki10Mexptimerunbook}
\end{subfigure} \hfil
\begin{subfigure}{.33\textwidth}
    \includegraphics[scale=.33]{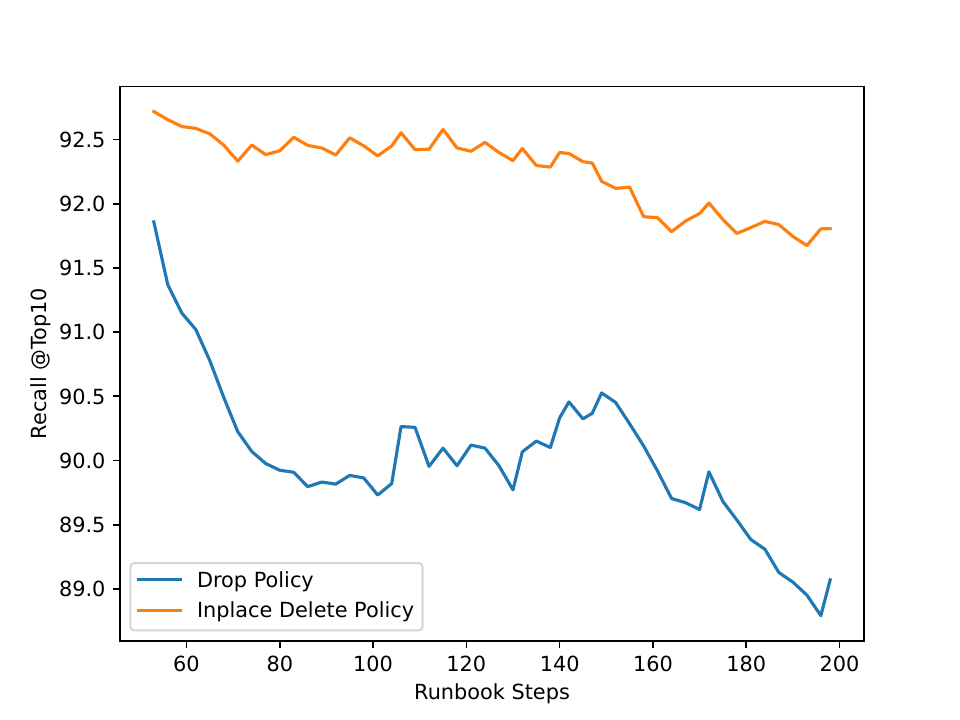}
    \caption{MS Turing-1M expiration time runbook. }
    \label{fig:msturingexptimerunbook}
\end{subfigure}\hfil
\begin{subfigure}{.33\textwidth}
    \includegraphics[scale=.33]{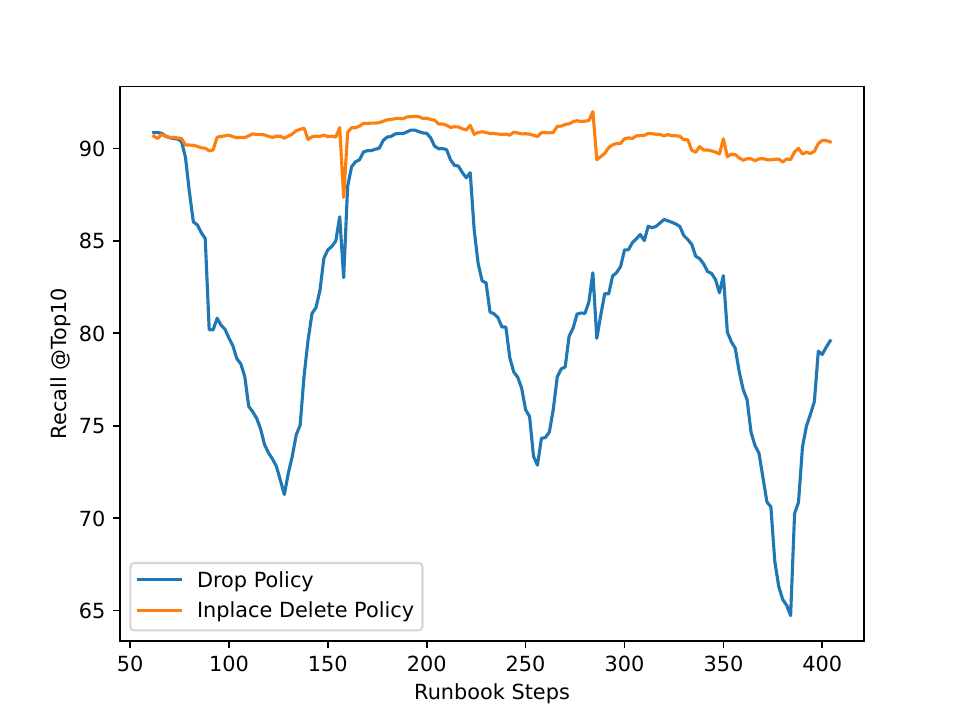}
    \caption{MSTuring-10M clustered runbook. }
    \label{fig:msturingclusteredrunbook}
\end{subfigure}
\vspace{-5pt}
\caption{Recall trends for the runbooks described in Section~\ref{sec:eval}. All runs used index parameters $R=32, L_{build}=100, L_{search}=200$. WikiCohere was compressed to 192 bytes and MSTuring to 50 bytes.}
\label{fig:runbooks}
\end{figure*}


We now consider the recall stability of a vector index in Cosmos DB using the runbooks described at the beginning of this section, plotted in Figures~\ref{fig:wiki10Mexptimerunbook} ,~\ref{fig:msturingexptimerunbook}, and~\ref{fig:msturingclusteredrunbook}. In all runbooks, an initial quantization is computed after 1000 vectors are inserted, and the index is re-quantized after 25000 vectors are inserted. The experiments compare the effects of taking no action other than dropping the deleted vector (labeled "Drop Policy" in the plots) versus using the inplace delete algorithm (Algorithm~\ref{alg:delete}). 

The experiments show that the in-place delete algorithm is critical for maintaining high recall under a stream of updates.
In the case of the expiration time runbooks (Figures~\ref{fig:wiki10Mexptimerunbook} and~\ref{fig:msturingexptimerunbook}), 
in-place delete increases average recall by 1-3 percentage points.
On the clustered runbook, which simulates distribution shift and is significantly more difficult than the expiration time runbooks,
in-place deletes achieve recall that is as much as 20 percentage points higher than the baseline,
meaning they are significantly more robust to distribution shifts. 

\subsection{Sharded indices for Multi-tenancy}
For multi-tenant apps, Cosmos DB allows user to create smaller DiskANN indices per tenant 
by  specifying a {\tt VectorIndexShardKey} in the indexing policy, 
one for each unique value of the selected document property.
This can significantly lower the query latency, cost and improve recall for queries that are focused on a tenant.

Table~\ref{tab:sharded} compares sharded DiskANN with a single DiskANN index on the YFCC dataset with year as the shard key.
Sharded DiskANN provides 3x lower mean and p99 latency with default query settings.
It also provided much higher recall at 98\%.
It was also more accurate than increasing search depth for the case without
sharding. The query RU charge also compared favorably.

\begin{table}[]
    \caption{Sharded Diskann with year as the shard key (row 1) compared with non-sharded DiskANN (rows 2, 3) on YFCC.
}
    \vspace{-10pt}
    \label{tab:sharded}
    \centering
    \begin{tabular}{c|c|c|c|c}
    \toprule
         & \multicolumn{3}{c|}{Latency (ms)} & \\ 
         Scenario &	Avg   &	P50  &	P99  &	Recall@10 \\ \midrule 
Sharded (Ls = 100) &	7.28 &	10.07 &	13.43 & 98.09 \\
Non-Sharded (Ls = 100) & 22.40 & 21.29 & 38.59 & 66.17 \\
Non-Sharded (Ls = 1000) & 94	& 95.47 &	164	& 87 \\
\bottomrule
    \end{tabular}

\end{table}

\section{Related Work}
\label{sec:related}

\subsection{Algorithms for Vector Search}
Algorithms for vector search fall roughly into two categories. The first category relies on partitioning the dataset into spatial cells and choosing a small number of cells to exhaustively search at query time. Common ways to partition include via clustering and locality-sensitive hashing (LSH). Some widely used partition-based vector search algorithms include FAISS~\cite{jegou2010product,douze2024faiss}, SPANN~\cite{SPANN}, ScaNN~\cite{ScaNN}, and many more~\cite{andoni2022falconn,YumingXuEtAl23}. Partition-based algorithms benefit from shorter indexing times than graph algorithms on average, but their query complexity increases much faster with the size of the dataset when compared to queries on graphs~\cite{ParlayANN24}.

Graph-based algorithms form a proximity graph with one vertex per embedding and greedily traverse the graph at query time to find a query's nearest neighbors. Examples include HNSW~\cite{HNSW16}, NSG~\cite{NSG17}, DiskANN~\cite{diskann19}, and ELPIS~\cite{elpis23}. Empirically, the query complexity of graphs scales logarithmically with the size of the dataset and is much better than LSH and clustering (see Figure~\ref{fig:graph_lsh_clustering}). Graph-based ANNS algorithms can maintain high recall in the streaming setting~\cite{freshdiskann, xu2025inplace}. Furthermore, they are adaptable to filtered queries~\cite{FilteredDiskANN} and queries that respect notions of diversity~\cite{anand2025diverse}. For these reasons---scalability, query efficiency, versatility, and robust updates---we use graphs in this work. 




\subsection{Vector Indices in Databases}
\emph{SingleStore-V}~\cite{singlestore-v}  supports both graph
and clustering based vector indices inside the SingleStore database.
The design loosely couples existing vector indexing libraries for HNSW and IVF algorithms.
SingleStore creates one vector index per segment, and rebuilds the indices as segments merge.
Each query has to fan out to indices over multiple segments rather than being served by one index.
More importantly, each vector index must be stored in memory for good performance,
which can be  expensive. For instance, a machine with 256GB memory and 32 cores is used for 10 million scale experiments. They  report using 3.8GB memory for the million sized  GIST1M dataset.
In contrast, we use <5GB of memory even for  10 million scale indices over 12 KB embeddings. 
Our system's ability to match query performance
with a small slice of machine's resources results in significantly lower system costs.

\emph{JVector}~\cite{JVector} is a Cassandra-based vector index with DiskANN.
They construct a DiskANN index in one shot and ingest it into the database, and  support incremental insertions and deletions using FreshDiskANN. Details on how the indices are managed and updated across partitions and segments are not documented to the best of our knowledge.
DataStax offers a managed product based on JVector. While their serverless offering does not provide an availability SLA, the enterprise provisioned capacity model has a 99.99\% SLA.
While their enterprise pricing is unspecified,
their standard tier has a monthly price of \$900/month
for 10 Queries/second over a million sized index.
The corresponding Cosmos DB monthly cost with Autoscale is less than \$50.

Elastic~\cite{elasticsearch-vector} offers vector search using a segment-based indexing system. 
It therefore has disadvantages similar to the SingleStore-V design. 
Furthermore, as a managed search engine, it does not offer the same level of robustness
and data consistency as as an operational database.
To compare costs, consider the 10 million vector index over Wiki-Cohere data.
Even with scalar quantization, we would need a machine with 15GB DRAM. 
We estimate using their \href{https://cloud.elastic.co/pricing}{pricing tool}
that the smallest such machine from Vector Search Optimized SKUs with 1.9vCPUS
would  costs about \$0.7/hour or about \$500/month.
A manually provisioned Cosmos DB service for 500 RU/sec or 5000 RU/sec
would serve 10 QPS and 100 QPS and cost about $\$30$ and $\$300$/month respectively.

pgvector~\cite{pgvector} integrates HNSW and IVF-Flat indexing algorithm inside PostgreSQL
and offers the inherent advantages of the PosgreSQL ecosystem.
Initial versions required machines large enough to fit the entire 
dataset as well as the index in memory for reasonable performance.
Scalar and binary quantizations have recently been added to reduce the memory requirement,
but performant indexing and updates still rely on the availability of much larger memory than our system.
pgvectorscale~\cite{pgvectorscale} similarly integrates DiskANN 
inside PostgreSQL and offers better performance than pgvector in limited memory settings.
We leave an exhaustive benchmarking against these systems for future work.
They primarily differ from our system in that they do not offer  flexible schemas,
multi-tenancy, and scale-out out of the box. 
V-Base~\cite{vbase} also builds on PostgreSQL and aims to improve vector search performance.
However it uses  an independent buffer pool  not integrated with
to PostgreSQL storage engines. Substantially more work is needed to build high-availability
and other features in this system. 

AnalyticDB-V~\cite{AnalyticDBV} pioneered the concept of integrated high-dimensional 
indices inside database systems that support hybrid queries with SQL semantics.
We do not directly compare with it since it is a provisioning based system.

\subsection{Specialized Serverless Vector DBs}
Specialized serverless vector databases are popular due to their ease-of-use,
flexible pay-per-use pricing structure, especially for early-stage applications,
or workloads with irregular traffic patterns.
However, these platforms currently exhibit limitations regarding enterprise-level readiness, particularly in terms of reliability, recovery, security, and compliance features.

For example, Pinecone~\cite{pinecone-enterprise-pricing} offers 99.95\% availability, but maintains only one data replica by default and provides limited backup-and-restore functionality applicable solely to the vector index, not the data itself. In terms of security, Pinecone supports Role-Based Access Control (RBAC) for both control and data plane operations without custom role support. Certifications include HIPAA BAA, AICPA SOC, GDPR, and ISO 27001 as of this writing.

Another offering, Zilliz~\cite{zilliz-enterprise-pricing}, provides no availability SLA, 
lacks replication and backup-and-restore,
and limits RBAC functionality to the control plane without custom roles.
Its compliance certifications are limited to SOC 2 Type II, ISO/IEC 27001, and GDPR.

Turbopuffer~\cite{turbopuffer} is another serverless vector database designed for multi-tenant usage.
However, it still lacks significant enterprise and robustness features such as RBAC, backup-and-restore,
global availability. Turbopuffer does not publicly list query price for 10 million vector shards, 
however, its query complexity increases substantially with the size of the tenant.
For example, cost per 1 million queries increases from \$3.58 to \$33.4 when the number
of documents per name space increases from 100K to a modest 1 million\cite{turbopuffer-pricing}.

In contrast, Azure Cosmos DB delivers robust enterprise readiness with a default availability
of 99.99\%, configurable up to 99.999\%, alongside four data replicas by default.
Distinct from any vector database solution, Cosmos DB has an SLA on document
reads and writes of 10ms for 1KB transactional documents.
It provides comprehensive RBAC capabilities for both control and data plane operations,
including customizable role definitions. Additionally, it is compliant on more than a dozen
standard including HIPAA BAA, FedRAMP, GDPR, ISO 27001, and others\cite{cosmosdbcompliance}.




\section{Conclusion}
We have designed a serverless, cost-efficient, and elastic
vector index in an operational database.
We were able to achieve this by using the existing indexing tree, document store,
and index manager in Cosmos DB, thus inheriting many of its structural advantages. 
To realize a highly efficient system, we redesigned the \diskann vector indexing library
from scratch to inter-operate with databases at sub-microsecond granularity.
As part of the redesign, we introduced support for handling asynchronous
requests and multiple logical indices from one \diskann instance.
We also introduced many algorithmic novelties to better align
with the databases. This design offers a general template for designing
vector indices in  databases.


\section{Acknowledgements}
We thank Philip A. Bernstein for his feedback on the work and suggestions for improving this draft.

\newpage



\bibliographystyle{ACM-Reference-Format}
\bibliography{ref}

\clearpage

\appendix
\section{DiskANN plots and algorithms}
\label{apdx:plotsalgs}

In this appendix we provide pseudocode for the core DiskANN algorithms described in the paper, as well as some supplementary plots on performance of the DiskANN library.

Algorithm~\ref{alg:greedysearch} shows the search algorithm on a DiskANN graph.  Algorithm~\ref{alg:insert} shows the insert procedure, while Algorithm~\ref{alg:robustprune} shows the pruning subroutine which is used in both insertion and deletion. Algorithm~\ref{alg:build} shows how the insert procedure is used to incrementally build the graph. Algorithm~\ref{alg:minibatchinsert} shows the generalization of the insertion routine to minibatch updates. Algorithm~\ref{alg:delete} specifies the consolidation algorithm that clears deleted vertices from the graph. Algorithm~\ref{alg:betasearch} shows the modification of greedy search to account for a query label $\mathcal{B}$.

\begin{algorithm}[t]
    \caption{Index Construction}
    \label{alg:build}
    \DontPrintSemicolon
    \KwData{Dataset $P$, degree bound $R$, list size parameter $L$, RNG parameter $\alpha$}
    \KwResult{Navigable Graph $G = (P,E)$, start node $s$}
    let $s \gets \arg \min_{p \in P} || \dbpoint{p} - \frac{\sum_{p' \in P} \dbpoint{p'}}{n} || $ \tcp{$s$ is the \emph{medoid} of the dataset}\;
    \For{each point $p$ in dataset $P$}{
           run $\insertpoint{\dbpoint{p}}{s}{L}{\alpha}{R}$
    }
\end{algorithm}

\begin{algorithm}[t]
\SetKwFor{ParallelFor}{parallel for}{do}{end parallel for}
  \DontPrintSemicolon
  \small
  \KwData{Graph $G(P,E)$ with start node $s$, set of new vectors $X$, parameter $\alpha > 1$, out degree bound $R$, list size $L$}
  \KwResult{Graph $G'(P',E')$ where $P' =  P \cup X$}
  \tt
  \ParallelFor{$x_p \in X$}{
      initialize expanded nodes $\cE \gets \emptyset$\;
      initialize candidate list $\mathcal{L} \gets \emptyset$\;
      $[\mathcal{L},\cE] \gets \greedysearch{s}{p}{1}{L}$        \;
      set $\nout{p} \gets \prune{p}{\cE}{\alpha}{R}$  \;
    }
    set $B \gets \bigcup_{x_p \in X} \nout{x_p}$ \;
    \ParallelFor{$b \in B$}{
        set $N \gets \{x_p | b \in \nout{x_p} \}$ \;
        set $N_b \gets N \cup \nout{b}$ \;
        set $\nout{b} \gets \prune{b}{N_b}{\alpha}{R}$ \;
    }
    \caption{\batchinsertpoint{X}{s}{L}{\alpha}{R}}
    \label{alg:minibatchinsert}
\end{algorithm}

\begin{algorithm}[t]
    \caption{Inplace Delete}
    \label{alg:delete}
    \DontPrintSemicolon
    \KwData{Dataset $P$, degree bound $R$, list size parameter $L$, RNG parameter $\alpha$, replace parameter $c$, deleted point $p \in P$}
    \KwResult{Navigable Graph $G = (P \setminus p)$, start node $s$}
    \Begin{
    $N_{twohop} \leftarrow N_{out}(N_{out}(p)) \cup N_{out}(p)$ \;
    $B \leftarrow \{b | b \in N_{twohop} \&\& p \in N_{out}(b)\}$ \;
    \For{$b \in B$}{
        $N_{out}(b) \leftarrow \{N_{out}(b) \setminus p\}$ \;
        $C \leftarrow c$ closest nodes to $b$ in $N_{out}(p)$ \;
        $N_{out}(b) \leftarrow N_{out}(b) \cup C$ \;
        \If{$|N_{out}(b)| > R$}{
            $N_{out}(b) \leftarrow$ RobustPrune($b, N_{out}(b), \alpha, R$) \;
        }
    }
    \For{$b \in N_{out}(p)$}{
        $C \leftarrow c$ closest nodes to $b$ in $N_{out}(p)$ \;
        \For{$c \in C$}{
            $N_{out}(c) \leftarrow N_{out}(c) \cup \{b\}$ \;
            \If{$|N_{out}(c)| > R$}{
                $N_{out}(c) \leftarrow$ RobustPrune($c, N_{out}(c), \alpha, R$) \;
            }
        }
    } 
     $N_{out}(p) \leftarrow \emptyset$ \;
     }
\end{algorithm}

\begin{algorithm}[t]
	\DontPrintSemicolon \small \KwData{ Graph $G$ with start node
          $s$, query \dbpoint{q}, label bitmap $\mathcal{B}$, result size $k$, search list size $L\geq k$, filter search parameter $0\leq \beta \leq 1$}
	
	\KwResult{Result set $\cL$ containing $k$-approx NNs with the same label bitmap as $x_q$, and a set $\cV$ containing all the visited nodes}

    \SetKwProg{Fn}{Function}{}{}
	
	\tt
	\Begin{
		initialize sets $\cL\gets \{s\}$, $\cE\gets\emptyset$, and $\cV\gets\emptyset$\; 
        \Fn{\texttt{dist($x_p$, $\mathcal{B}_p$, $x_q$, $\mathcal{B}_q$)}}{
            $d \gets ||x_p-x_q||$ \;
            \If{$\mathcal{B}_p == \mathcal{B}_q$}{
                return $\beta * d$ \;
            } \Else{
                return $d$ \;
            }
        }
		\While{$\cL \setminus \cE \neq \emptyset$}{
			let $p* \gets \arg \min_{p \in \cL\setminus \cE} \text{dist}( \dbpoint{p}, \mathcal{B}, \dbpoint{q}, \dbpoint{q}.\text{label})$\;
			update $\cL \gets \cL \cup (\nout{p^*} \setminus \cV)$ and $\cE \gets \cE \cup \{p^*\}$\;
			\If{$|\cL| > L$} {
				update $\cL$ to retain closest $L$ points to \dbpoint{q}\;
			}
                update $\cV \gets \cV \cup \nout{p^*}$
		}
		return $[$closest $k$ points from $\cV$ with label $\mathcal{B}$; $\cV$$]$\;
	}
	\caption{\betasearch{s}{\dbpoint{q}}{\mathcal{B}}{k}{L}{\beta}}
	\label{alg:betasearch}
\end{algorithm}

Next, we provide some experimental evidence supporting the conclusions in this paper on the efficiency of graph algorithms over partition-based algorithms. 

\begin{figure}
    \centering
    \includegraphics[width=\linewidth]{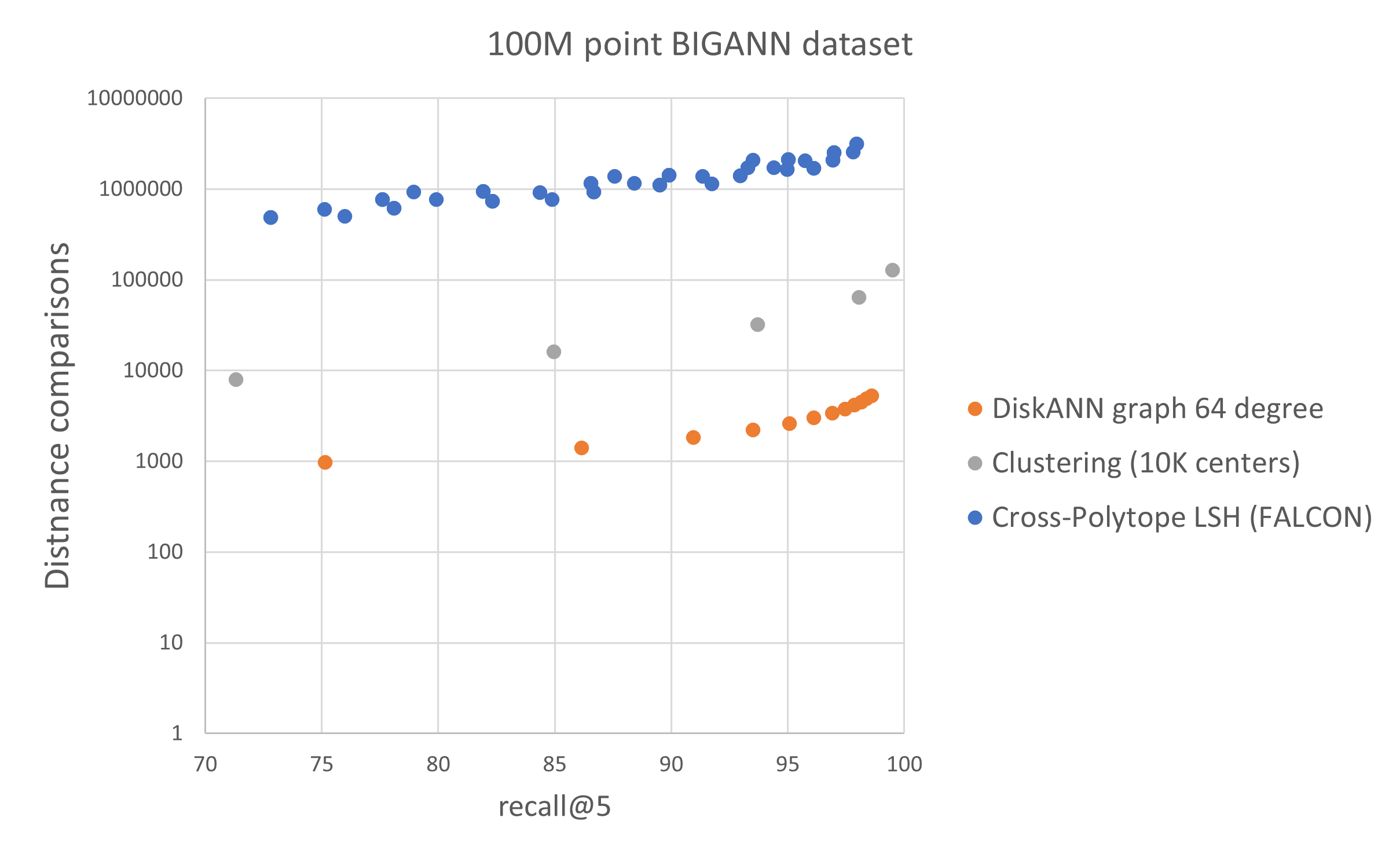}
    \caption{Comparison of number of distance comparions needed for different values of recall@5
    between DiskANN, Clustering (10,000) centers and the Cross-polytope LSH from FALCONN
    library [Razenshteyn’16]. Parameters for LSH are swept from number of tables = \{10, 15\},
    probe width = \{25 50 100 200\}, dimension of last polytope = \{4 8 16 32\},
    cross polytope number = 4, number of rotations = 3.}
    \label{fig:graph_lsh_clustering}
\end{figure}

\section{\cosmosdb and DiskANN interoperability }
\label{sec:interop}

\begin{figure}
    \centering
    \includegraphics[width=\linewidth]{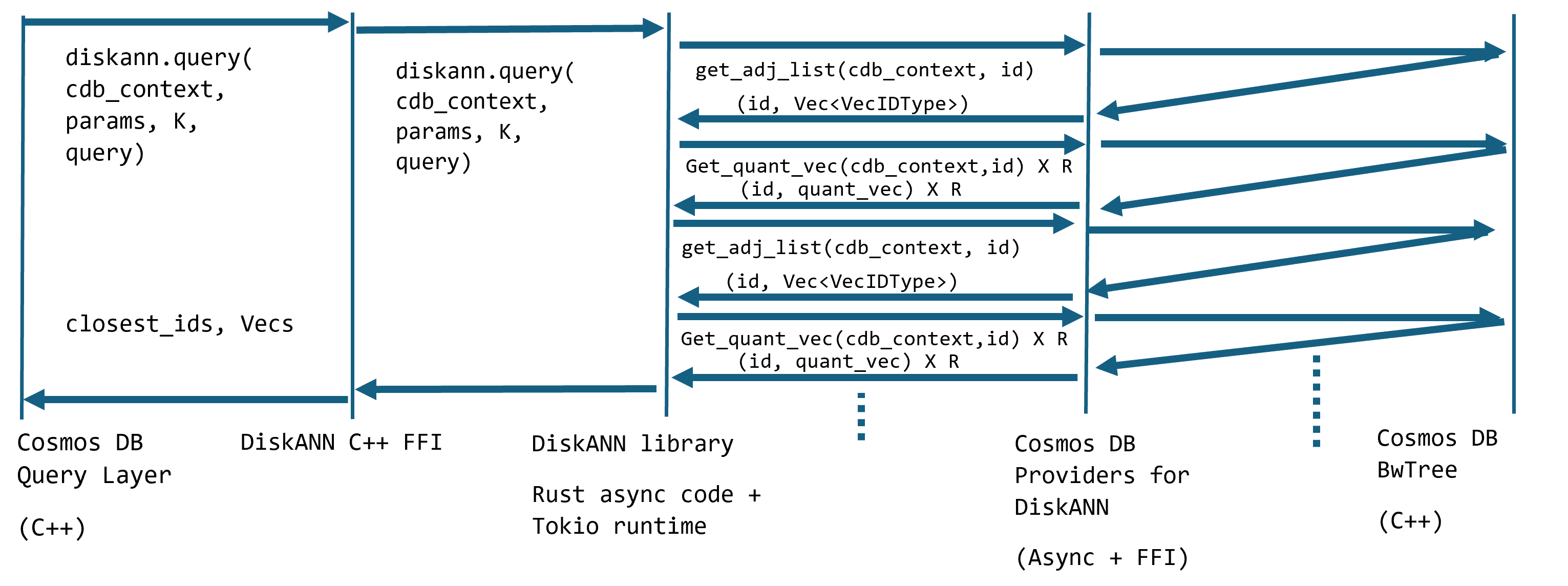}
    \vspace{-10pt}
    \caption{Control flow of a DiskANN query in \cosmosdb.}
    \label{fig:control-flow}
\end{figure}

To insert, the \cosmosdb IndexManager calls
{\tt diskann.insert} with appropriate execution context, which
in turn accesses the Bw-Tree index and the Document Store.
The DiskANN library written in Rust is sandwiched between IndexManager and Bw-Tree/Document Store  
components written in C++.
The interleaving is similar for the query path, and is illustrated in Figure~\ref{fig:control-flow}.  Since all of these components are written in an asynchronous programming style with different languages and run times,
this leads to an interesting challenge of implementing asynchronous 
callbacks across language  boundary.
The standard way to integrate Rust/C++ using C ABI  has no native support for async calls,
and public FFI libraries that support C++/Rust async inter-op are non-existent.
We build an internal solution to address this, 
and did not observe any significant overheads from this interface.

Asynchronous Rust uses cooperative tasks which are polled to completion.
When polling, a Rust asynchronous runtime provides tasks with a ``waker" that moves a task from an ``idle" state to a ``ready to run" state.
A task that cannot make progress when polled must store this waker and set up some mechanism by which the waker is invoked when progress can be made.
The runtime then responds to waker invocations by polling the task again.
Note that runtimes may also poll suspended tasks eagerly rather than waiting for a waker invocation.

Our asynchronous interface involves allocating shared data structures on the Rust side and providing C++ exclusive mutable access to this shared data.
The C++ side then releases it's exclusive access back to Rust, allowing Rust to make forward progress.

In more detail, to share an object $x$ (e.g., a data for a quantized vector) asynchronously, we first run Algorithm~\ref{alg:setup_async_call}.
This embeds $x$ inside a reference counted data structure, leaks one reference count to C++, and shared the embedded $x$ with C++.
At this point, C++ has complete ownership of the shared object $x$ and can perform backend operations required to populate $x$ with the correct data.
Reference counting ensures that C++ still has a valid object to work with if the Rust side tasks gets canceled.

\begin{algorithm}[t]
    \caption{Asynchronous Set Up (Rust)}
    \label{alg:setup_async_call}
    \DontPrintSemicolon
    \KwData{Data structure $x$ to share with C++}
    \KwResult{Reference counted data structure $A$ for synchronization}
    
    \tt
    \Begin{
        // Create reference counted data structure $A$ containing $x$, an atomic boolean $done$, and a Rust $waker$ protected by a Mutex
        
        // Increment and leak reference count for $A$
        
        // Provide C++ with a mutable pointer to $x$ embedded in $A$, a constant pointer to $A$, and a function pointer to Algorithm \ref{alg:asynchronous_callback}
        
        \Return{$A$}
    }
\end{algorithm}

Ownership of $x$ is released by C++ by Algorithm~\ref{alg:asynchronous_callback}.
This operation atomically synchronizes the shared state in $x$, invokes the waker is present, and decrements the reference count previously leaked to C++.
After C++ makes this callback, it is not allowed to access the shared object $x$ again.

\begin{algorithm}[t]
    \caption{Asynchronous Callback (C++)}
    \label{alg:asynchronous_callback}
    \DontPrintSemicolon
    \KwData{A pointer $pA$ to object $A$ from Algorithm \ref{alg:setup_async_call}}
    
    \tt
    \Begin{
        $A = \text{restore\_from\_pointer}(pA)$
        
        // This releases C++ ownership of $x$
        
        $A.done.store(true,\,\text{memory ordering release})$\;
        
        // Lock and if present, invoke $A.waker$ 
        
        // Implicitly run the destructor for $A$, decrementing the reference count
    }
\end{algorithm}

The final piece of this interface is the Rust side polling in Algorithm~\ref{alg:asynchronous_poll}.
The algorithm begins by setting the runtime's waker in the shared state, then atomically checking if the callback has been completed.
If the callback is complete, Rust reassumes ownership of $x$ and makes forward progress.
Otherwise, the task becomes idle and waits for the callback to be made.
Note that the waker inside the reference counted object is stored behind a mutex to ensure that completion of the callback is always observed by Rust.

\begin{algorithm}[t]
    \caption{Asynchronous Poll (Rust)}
    \label{alg:asynchronous_poll}
    \DontPrintSemicolon
    \KwData{Object $A$ from Algorithm \ref{alg:setup_async_call} and a Rust $waker$}
    \KwResult{$Pending$ if Algorithm \ref{alg:asynchronous_callback} has not been made, otherwise the modified $x$}
    
    \tt
    \Begin{
        // The below operation is done under a Mutex
        
        $A.waker = waker$

        // Check if the callback has been made
        
        $isdone = A.done.load(\text{memory ordering acquire})$

        \eIf{isdone}{
            \Return{$A.x$}
        }{
            \Return{$Pending$}
        }
    }
\end{algorithm}

The asynchronous protocol can be naturally extended to support bulk operations by allowing the asynchronous callback in Algorithm~\ref{alg:asynchronous_callback} to release just part of the shared state in $x$ with suitable modifications to Algorithm~\ref{alg:asynchronous_poll}.
This can be used to group all quantized vector reads for an expansion step in Algorithm~\ref{alg:greedysearch} into a single FFI setup call, reducing overhead.

\paragraph{Resource Governance}
\cosmosdb uses co-operative scheduling where all operations 
belonging to multiple collections frequently check with
the resource governor for available budget
to continue executing. When their resource is exhausted,
they temporarily yield control and resume later.
Further a logical activity (unit of work) in \cosmosdb backend can be scheduled to execute across multiple threads during its lifetime.

To track the resource consumption of Rust-side operations, each spawned task is embedded inside  a monitoring layer.
When the \texttt{tokio} runtime polls a task, the monitoring layer begins CPU profiling before polling the embedded task.
When the embedded tasks exits polling, whether via completion or to cooperatively yield, the monitoring layer accumulates the CPU resources (thread cycles) utilized.
Because asynchronous Rust tasks can only migrate between threads when idle (i.e., not being polled), this technique provides a low-overhead method of measuring compute resources used directly for an activity.

\cosmosdb can use this resource consumption to temporarily suspend a Rust-side task at the asynchronous FFI layer when the task's associated logical activity exceeds its budget quota.

\section{IndexTerm Examples}
~\label{sec:indexterm-example}
Here is an example of a JSON document with a '/embedding' path.
\begin{verbatim}
    {
      "embedding": [
            -0.029442504048347473,
            -0.06647671014070511,
            -0.22546832263469696,
            ...
        ]
    }
\end{verbatim}

\textbf{Inverted Term}

Inverted term design for Quantizated data.
\begin{align*}
\text{TermKey} =\ & 
\underbrace{\texttt{0xD278B10FF01E964D33CC55AA01FE28B9}}_{\text{'/embedding' Prefix (16 Bytes)}} \\
& +\ \underbrace{\texttt{0x17}}_{\text{'QuantTerm' Type (1 Byte)}} \\
& +\ \underbrace{\texttt{0xAB12C3D4E5F60718|0x67..}}_{\text{Document Id + QuantizedVector Value (Variable Length)}}
\end{align*}

\begin{align*}
\text{TermValue} = 
\underbrace{\texttt{Dummy Bitmap (for all values, irrespective of keys)}}_{\text{PES Bitmap}}
\end{align*}

\textbf{Forward Term Representation}. 

Forward term design for Adjacency List data.
\begin{align*}
\text{TermKey} =\ & 
\underbrace{\texttt{0xD278B10FF01E964D33CC55AA01FE28B9}}_{\text{'/embedding' Prefix (16 Bytes)}} \\
& +\ \underbrace{\texttt{0x18}}_{\text{'AdjacencyTerm' Type (1 Byte)}} \\
& +\ \underbrace{\texttt{0xAB12C3D4E5F60718}}_{\text{DocumentId (8 Bytes)}}
\end{align*}

\begin{align*}
\text{TermValue} =\ & 
\underbrace{\texttt{[(0xDE4203A985225529), (DocId2), (DocIdN), ....]}}_{\text{Variable Length Adjacency List (8 Byte DocumentIds)}}
\end{align*}

\section{VectorDistance function}
\label{sec:vec-dist}

The VectorDistance system function takes in the following arguments:

\begin{table}[H]
\centering
\begin{tabular}{| l | p{5cm} |}
\hline
\textbf{Parameter} & \textbf{Description} \\ \hline
vector\_expr\_1 & This argument can be an array literal or a path to a property in the document. \\ \hline
vector\_expr\_2 & This argument can be an array literal or a path to a property in the document. \\ \hline
bool\_expr & This argument indicates if the search needs to be an exact search. “true” indicates that it needs to be an exact search so the execution plan will not use any index that provides approximate results \\ \hline
obj\_expr & An optional JSON object that is used to specify options that can control the vector distance calculation. Some of the options are: distanceFunction,  datatype, searchListSizeMultipier, quantizedVectorListMultiplier.\\ \hline
\end{tabular}
\end{table}

\section{Ingesting popular datasets on \cosmosdb}
An open source tool \cite{VectorIndexScenario-Suite} was released by the team to one-click ingest and query popular open datasets with \cosmosdb. Extending the tool to ingest custom datasets is relatively simple and customers can use it experiment with Vector Indexing and \cosmosdb. 

\section{Additional query latency and RU charge plots}
\label{sec:apdx-query-plots}

\begin{figure}
    \centering
    \includegraphics[width=0.9\linewidth]{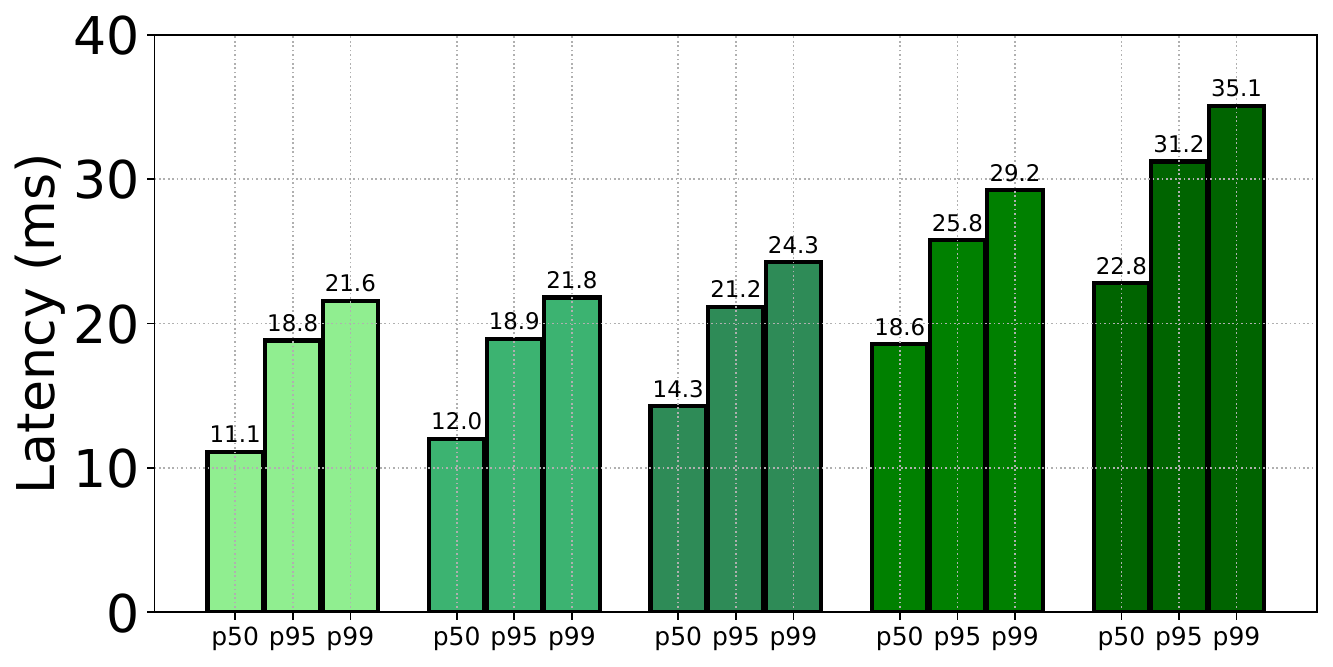}
    \includegraphics[width=0.9\linewidth]{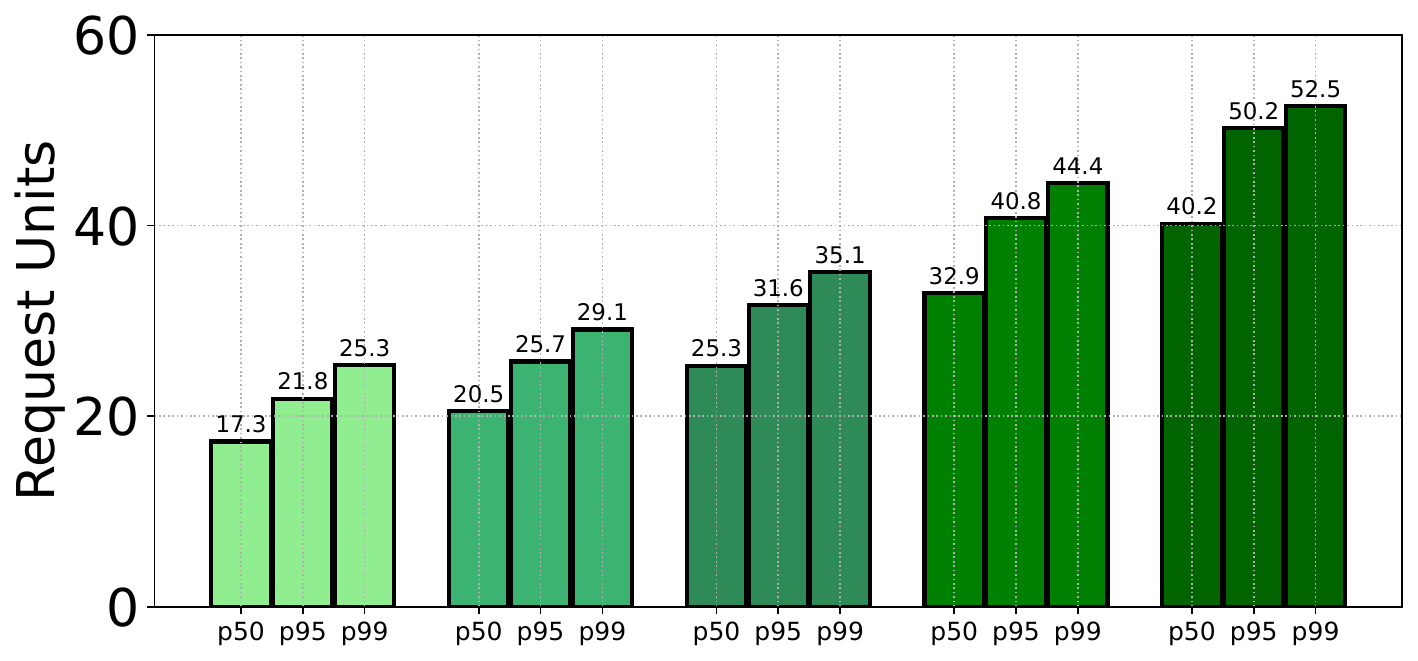}
    \includegraphics[width=\linewidth]{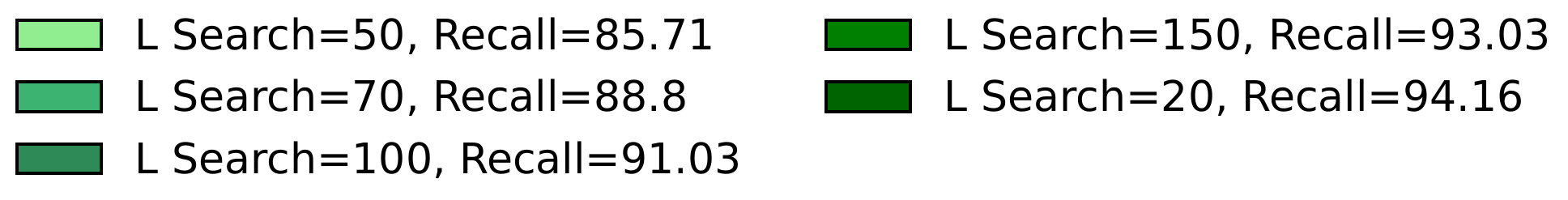}
    \caption{p50, p95 and p99 query latencies and query RU charge for 1 million Wiki-Cohere vector index for various values
    of search list size, and the corresponding recall@10.}
    \label{fig:wiki-1M-query-latency}
\end{figure}

\begin{figure}
    \centering
    \includegraphics[width=0.9\linewidth]{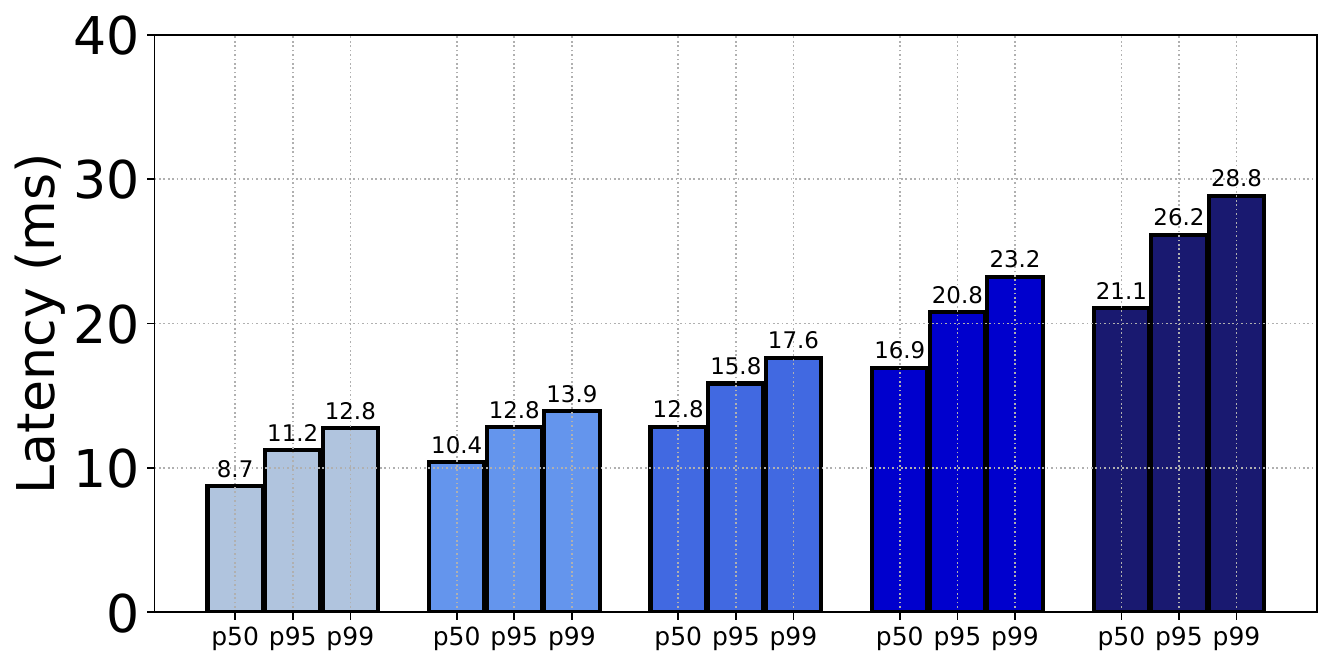}
    \includegraphics[width=0.9\linewidth]{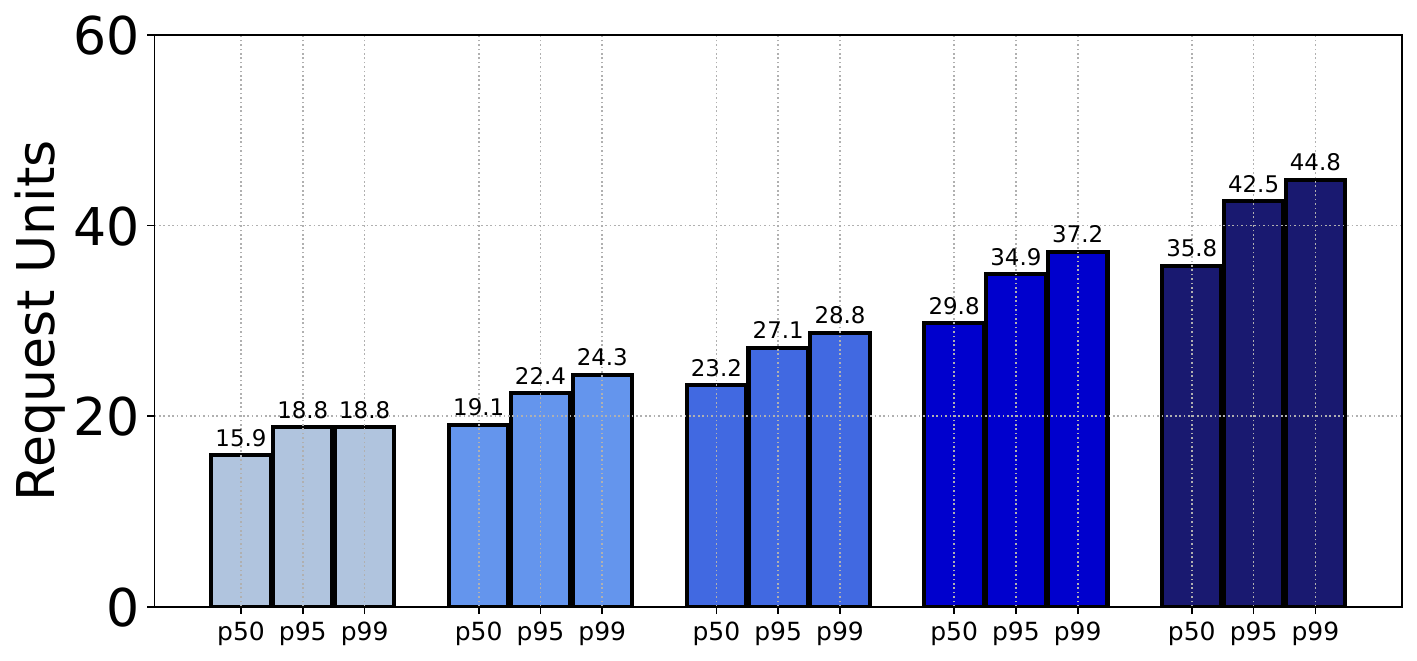}
    \includegraphics[width=\linewidth]{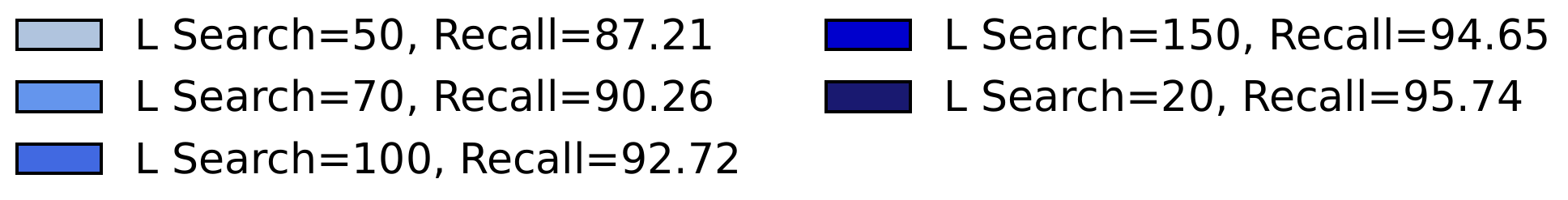}
    \caption{p50, p95 and p99 query latencies and query RU charge for 100,000 Wiki-Cohere vector index for various values
    of search list size, and the corresponding recall@10.}
    \label{fig:wiki-100K-query-latency}
\end{figure}

Figures~\ref{fig:wiki-1M-query-latency} and~\ref{fig:wiki-100K-query-latency}
present additional data to track the relation
between search parameter, recall and query latency and RU cost.

\end{document}